\documentclass[aps,amsmath,amssymb,superscriptaddress,floatfix,showpacs,letterpaper,nofootinbib]{revtex4}
\pdfoutput=1

\usepackage{graphicx}
\usepackage{enumerate}

\begin{document}

\title{Grand Unification Through Gravitational Effects}

\author{Xavier Calmet}\email{x.calmet@sussex.ac.uk}\affiliation{Physics and Astronomy, University of Sussex, Falmer, Brighton, BN1 9QH, UK}
\author{Stephen D.~H.~Hsu}\email{hsu@uoregon.edu}\affiliation{Institute of Theoretical Science, University of Oregon, Eugene, OR 97403, USA}
\author{David Reeb}\email{dreeb@uoregon.edu}\affiliation{Institute of Theoretical Science, University of Oregon, Eugene, OR 97403, USA}

\begin{abstract}
We systematically study the unification of gauge couplings in the presence of (one or more) effective dimension-5 operators $cHG_{\mu\nu}G^{\mu\nu}/4M_{Pl}$, induced into the grand unified theory by gravitational interactions at the Planck scale. These operators alter the usual condition for gauge coupling unification, which can, depending on the Higgs content $H$ and vacuum expectation value, result in unification at scales $M_X$ significantly different than naively expected. We find non-supersymmetric models of $SU(5)$ and $SO(10)$ unification, with natural Wilson coefficients $c$, that easily satisfy the constraints from proton decay. Furthermore, gauge coupling unification at scales as high as the Planck scale seems feasible, possibly hinting at simultaneous unification of gauge and gravitational interactions. In the Appendix we work out the group theoretical aspects of this scenario for $SU(5)$ and $SO(10)$ unified groups in detail; this material is also relevant in the analysis of non-universal gaugino masses obtained from supergravity.
\end{abstract}

\date{February 2010}


\maketitle

\section{Introduction and outline}\label{introductionsection}
What are the boundary conditions for grand unification? One typically assumes that the gauge couplings of the broken subgroups must become numerically equal at the unification scale $M_X$ \cite{Georgi:1974yf}. However, effects from physics above the unification scale can alter the gauge coupling unification condition. In an effective field theory approach, such effects can be caused by dimension-5 operators of the form $cHG_{\mu\nu}G^{\mu\nu}/4M_{Pl}$, which shift the coefficients of the gauge kinetic terms in the low-energy theory after the Higgs $H$ acquires a vacuum expectation value in grand unified symmetry breaking \cite{Hill:1983xh,Shafi}; one obvious source of such operators is quantum gravitational effects. Indeed, it would be {\it unnatural} (or require some special explanation) to assume that the Wilson coefficients $c$ above be zero or especially small \cite{tHooft:1979bh}; the default assumption should be that these coefficients are of order unity in grand unified models, with consequent unification conditions.

In conventional unification models, one might expect $\langle H \rangle \sim 10^{16}$ GeV, plausibly leading to effects from quantum gravity of order a fraction of a percent, $\langle H\rangle/M_{Pl}\sim10^{-3}$, on the gauge coupling unification condition. In \cite{ourprl} we showed that these dimension-5 operators can be even more relevant than previously suspected since the Planck mass $M_{Pl}$ tends to be smaller than naively assumed due to its renormalization group evolution \cite{Larsen:1995ax,Calmet:2008tn} under the influence of the large number of fields in supersymmetric grand unified theories. It was noted \cite{ourprl} that these dimension-5 operators introduce in supersymmetric unification models an uncertainty that can be bigger than the two-loop effects which are considered to be necessary to obtain good numerical unification of the gauge couplings.

The aim of this paper is different. We study whether the dimension-5 operators discussed above can lead to perfect gauge coupling unification {\it without} supersymmetry by their modifying of the gauge coupling unification condition. This unification scheme has been studied previously in the literature for models with and without supersymmetry, e.g.~in \cite{Rizzo:1984mk,Hill:1983xh,Shafi,Hall:1992kq,Vayonakis:1993nn,Datta:1995as,Dasgupta:1995js,Huitu:1999eh,Tobe:2003yj,ourprl,Chakrabortty:2008zk}, but in less detail and generality and mostly only the effect from a single gravitational operator has been considered.

In particular, in this paper we examine, in a systematic way, the effects when two or more dimension-5 operators are present in a theory. Unification under multiple dimension-5 operators has been studied before \cite{Huitu:1999eh} for the supersymmetric case \cite{susyguts}, for which, however, viable gauge coupling unification is well known \cite{Jones:1982,Marciano:1981,Amaldi:1991cn}. The main result of the present paper is that the measurement of the gauge couplings at the $Z$ mass \cite{Amsler:2008zzb} is compatible with non-supersymmetric grand unification based on $SU(5)$ or bigger groups like $SO(10)$. That is, grand unification does not require a supersymmetric extension of the standard model for a range of natural values of the Wilson coefficients $c$.

As opposed to models with one dimension-5 operator, in unified models with two or more such operators, the unification scale $M_X$ can be varied in a continuous and controlled manner with the Wilson coefficients $c$ and Higgs vacuum expectation values (vevs), and we examine this quantitatively in $SU(5)$ and $SO(10)$ theories, primarily for the non-supersymmetric case (cf.~\cite{Huitu:1999eh} for the supersymmetric case). We find that unification under this scheme can naturally raise the unification scale $M_X$ much above the conventional $10^{15}\,\text{GeV}$ to $10^{16}\,\text{GeV}$, and even allows to obtain gauge coupling unification at or near the Planck mass which is suggestive of unification of all gauge and gravitational forces at a common scale (see also \cite{Agashe:2002pr,MPlUnification}). Any such unification is safe from the proton decay constraint. This, and the fact that exact gauge coupling unification in the first place can be accomplished, may be a useful tool for model building.

Note, due to the gravitational origin of these dimension-5 operators, the unification scheme considered here is distinct from the Lavoura-Wolfenstein result \cite{Lavoura:1993su} or similar effects \cite{Weinberg:1980wa}, according to which non-supersymmetric unification can result from particle thresholds. In particular, as can be seen from the form of these effective operators, their effect increases as the unification scale gets closer to the Planck scale, i.e., as the Higgs vevs get bigger, which in turn allows to self-consistently shift $M_X$ to values much larger than conventionally assumed.

Realistic supersymmetric unification theories have issues linked to their large particle content. One is the presence of Landau poles between the unification and the Planck scale, another is possible lack of calculability in these theories due to potentially large running of the Planck mass \cite{ourprl}. Both issues are avoided in their non-supersymmetric counterparts. On the other hand, if supersymmetry is abandoned, the unification scale needs to be stabilized with respect to the gravitational scale (as, of course, does the weak scale). This ``little hierarchy problem'' can be avoided by one of the scenarios mentioned above where grand unification happens close to the Planck mass, thereby reducing or eliminating the hierarchy between the two scales.

\bigskip

\paragraph*{Outline of the paper.} For a physical description of the scenario and effects and for exemplary numerical results, see the Introduction and Summary, Sections \ref{introductionsection} and \ref{conclusionsection}. More detail is given in the main text, and the group theoretical formalism in the Appendix.

In Section \ref{sectionSU5setup}, we introduce the dimension-5 operators $cHG_{\mu\nu}G^{\mu\nu}/4M_{Pl}$ under investigation, first specifically for the case of an $SU(5)$ grand unified theory, and also discuss possible sensible choices for the Planck scale $M_{Pl}$ suppressing these operators. We then describe how these operators modify the usual gauge coupling unification condition. This section sets up the notation necessary to understand the figures, tables and most of the details in the main text.

In Section \ref{sectionSU5results}, we look at numerical results for possible gauge coupling unification under the modified gauge coupling unification condition in non-supersymmetric $SU(5)$ models. First (Section \ref{SU5oneoperatorsection}) we review the known effects in the presence of a single such operator \cite{Hill:1983xh,Shafi,Hall:1992kq,Chakrabortty:2008zk}, also addressing uncertainties in low-energy measurements of the gauge couplings and modifications due to two-loop running. Then, in Section \ref{twoHiggsSU5subsection}, it is numerically shown that (and how) the unification scale may be \emph{continuously} varied in models with two different dimension-5 operators. This, being one of the main results of this paper, is also true in models with more such operators, and we give the general treatment in Section \ref{generalmodelssection}, also deriving an estimate to assess which unification scales $M_X$ are achievable naturally.

With these numerical results and estimates, in Section \ref{protondecaysection} we see that even the lowest unification scales $M_X$, that are achievable in a natural way through our effect, are not in conflict with the current lower bounds on the proton lifetime, coming from the non-observation of proton decay so far. Thus, the dimension-5 operators can facilitate exact gauge coupling unification without supersymmetry and also allow the proton decay limit to be evaded (unlike in naive $SU(5)$ unification). We also investigate how further improvements in the proton lifetime bound constrain our models.

In Section \ref{gaugegravitysection}, we note that numerical gauge coupling unification is possible near or at the Planck scale in a very natural way in models with two or more dimension-5 operators. We speculate that this might hint at unification of the gauge interactions and gravity at a common scale (``gauge-gravity unification'', see \cite{Agashe:2002pr}).

Section \ref{sectionSO10} contains the scenario for an $SO(10)$ grand unified gauge group. The setup and unification results for exemplary cases are described, analogous to the $SU(5)$ case in the preceding sections. The main differences to $SU(5)$ are emphasized, namely the fact that a continuously varying $M_X$ can be achieved with merely a single dimension-5 operator, and the possible role of these operators in $SO(10)$ breaking with intermediate scales is briefly described.

In Section \ref{susysection} we briefly look at the effect of the dimension-5 operators in models of supersymmetric unification \cite{susyguts,Jones:1982,Marciano:1981}, making contact to the literature (e.g., \cite{Huitu:1999eh,ourprl,Chakrabortty:2008zk}). We find that in supersymmetric models the unification scale can also be shifted around, although unification close to the expected $2\times10^{16}\,\text{GeV}$ seems most likely. The main part of the present paper focuses on non-supersymmetric models as it is those where viable unification is commonly believed to be difficult or impossible (for further reasons, see the last paragraph of Section \ref{sectionSU5setup}).

We conclude and summarize the main ideas and mechanisms in Section \ref{conclusionsection}, illustrated with some numerical examples showing the size of the effects.

In an extensive Appendix, we present all normalization conventions and group theoretical details. In particular we carefully derive and give all of the Clebsch-Gordan coefficients $\Phi^s_{(r)t}$ associated with $SU(5)$ and $SO(10)$ breaking (the latter case takes up most of the work) to the standard model, in different bases $\{t\}$; only the coefficients relating to the standard model gauge group factors $s$ have been given before in the literature, see especially \cite{Martin:2009ad}, and often only their relative sizes, whereas we here follow a uniform absolute normalization scheme across different representations $r$. These analytical results are also useful for the treatment of non-universal gaugino masses obtained from ${\cal N}=1$ supergravity models of $SU(5)$ or $SO(10)$, for which the group theory involved is very similar, see, e.g., \cite{Ellis:1985jn,Drees:1985bx}. The Appendix furthermore describes the modified gauge coupling unification condition and a systematic method (which was used for the numerical work in this paper) for its solution.

\section{Dimension-5 operators and unification condition -- ${\bf{SU(5)}}$ case}\label{sectionSU5setup}
In this section we describe, specifically for the case of non-supersymmetric $SU(5)$ grand unification models \cite{GeorgiGlashow}, first the operators necessary for our scenario and then their effect on the condition for gauge coupling unification. The $SO(10)$ case holds additional subtleties and possibilities, to be deferred to Section \ref{sectionSO10}. We present actual numerical unification results for the $SU(5)$ case in Section \ref{sectionSU5results} and elaborate on physically interesting scenarios in Sections \ref{protondecaysection} and \ref{gaugegravitysection}.

At the energies available to us in present-day particle physics experiments, Nature is described very well by non-Abelian gauge field theories which are non-gravitational, although we know that at some energy scale a more complete description must take gravity into account. In other words, at our energies the effective Lagrangian of the world is a gauge theory with certain additional non-renormalizable operators of dimension 5 and higher, accounting for the gravitational interactions that have been integrated out. These operators must obey the symmetries (gauge and Lorentz invariance, etc.) of the low-energy theory and are suppressed by powers of the gravitational scale $M_{Pl}$, so they are seemingly negligible at our energies. By this logic, grand unification also appears as an effective theory, valid between the unification scale $M_X$ and well below the Planck scale $M_{Pl}$, and contains higher-dimensional operators induced by gravity and suppressed by $M_{Pl}$; at energies $\sim M_X$, however, such operators are potentially much more significant due to the proximity of the scales $M_X$ and $M_{Pl}$.

One set of such dimension-5 operators, that may have important effects in grand unification, are singlets formed from gauge field strengths $G_{\mu\nu}$ and Higgs multiplets $H_i$ of the grand unified gauge group $G=SU(5)$,
\begin{equation}\label{dim5operators}
{\cal L}=\frac{c_i}{4M_{Pl}}H_i^{ab}G^a_{\mu\nu}G^{b\mu\nu}~,
\end{equation}
suppressed by one power of the Planck mass $M_{Pl}$ such that $c_i$ are dimensionless (Wilson) coefficients. The index $i$ is summed implicitly and includes the possibility that the effective Lagrangian may contain several such operators involving different Higgs multiplets $H_i$ of the theory, which will be one of our main tools later on. In an $SU(5)$ gauge theory, operators (\ref{dim5operators}) can be formed gauge-invariantly only with $H_i$ in the representations $r_i={\bf 1}$, ${\bf 24}$, ${\bf 75}$ or ${\bf 200}$ (although the theory might contain additional multiplets in other representations); these irreducible representations (irreps) can all uniformly be written in component notation with symmetric adjoint indices $a$, $b$ of the gauge group, which establishes a \emph{common} normalization for the different operators $i$ in (\ref{dim5operators}). (Here and later, see the Appendix for a careful treatment of the relevant group theoretical aspects and normalization conventions; see also \cite{Slansky:1981yr}.)

The dimension-5 operators (\ref{dim5operators}) are suppressed by the Planck scale $M_{Pl}$, the energy scale at which quantum gravity sets in, which, as an interaction not accounted for by the renormalizable terms in the Lagrangian, induces these effective operators. There is some ambiguity (or arbitrariness) as to whether the appropriate suppression scale $M_{Pl}$ should be taken to be the ``naive'' Planck scale $G_N^{-1/2}=1.2\times10^{19}\,\text{GeV}$ or, more commonly, the reduced Planck scale $\left(8\pi G_N\right)^{-1/2}=2.4\times10^{18}\,\text{GeV}$, as this is the quantity that controls quantum gravity computations. To leave this choice open, we parametrize
\begin{equation}\label{suppressionscale}
M_{Pl}\equiv\frac{G_N^{-1/2}}{\xi} = \frac{1.2\times10^{19}\,\text{GeV}}{\xi}~;
\end{equation}
$\xi=1$ corresponds to the choice of the naive Planck scale as the suppression scale, $\xi_\text{red}=\sqrt{8\pi}\approx5$ to the reduced Planck scale, which is what we assume (implicitly) in most discussions. The smaller a suppression scale $M_{Pl}$ one accepts, i.e.~the bigger $\xi$ one chooses, the more pronounced the effects of the operators (\ref{dim5operators}) will be, at fixed Wilson coefficients $c_i$. Effects of equal size can be achieved for indirectly proportional coefficients $c_i\to c_i/\xi$ when changing $\xi$. (Conservatively, we put an explicit factor $1/4$ in (\ref{dim5operators}) to avoid overcounting of terms in the contraction of two gauge field strengths, just as in the canonical gauge boson kinetic term.)

Also concerning the choice of an appropriate suppression scale $M_{Pl}$, we have shown previously \cite{Calmet:2008tn,ourprl} that the fundamental value of Newton's constant (i.e., at high energies) is different from its observed low-energy value $G_N$ used in (\ref{suppressionscale}): matter field fluctuations of $N_0$ real scalar, $N_{1/2}$ Weyl fermion and $N_1$ gauge boson fields lead to a running of Newton's constant
\begin{equation}\label{runningGN}
\frac{1}{G(\mu)}=\frac{1}{G_N}-\frac{\mu^2}{12\pi}\left(N_0+N_{1/2}-4N_1\right)
\end{equation}
at one loop, similar to the running of gauge couplings (see also \cite{Kabat:1995eq,Larsen:1995ax,Percacci:2005wu}). Then determining the fundamental gravitational scale $M_{Pl}$ via $G(\mu=M_{Pl})^{-1/2}\sim M_{Pl}$ yields a value lower by a factor of $\eta\equiv\sqrt{1+(N_0+N_{1/2}-4N_1)/12\pi}$ than without the running. This additional change in the suppression scale $M_{Pl}$ is easily incorporated into our parametrization (\ref{suppressionscale}) via $\xi\to\xi^\text{run}=\xi\eta$, and Table \ref{etatable} illustrates the size of this running effect in various grand unified models. Furthermore, as illustrated in the last row of the table, appropriate choices of $\xi$ can accommodate suppression scales $M_{Pl}$ of other origin as well, e.g.~string compactification scales as in \cite{Shafi}.

\begin{table}[htb]
\begin{center}
\begin{tabular}{l||c|c|c||c}
\hline
particle physics model & $N=N_0+N_{1/2}-4N_1$ & $\eta=\sqrt{1+N/12\pi}=\xi^\text{run}$ & $\xi^\text{run}_\text{red}=\sqrt{8\pi}\eta$ & $\log_{10}M_{np}/M_X$ \\
\hline
\hline
no running of $G_N$ & & $1$ & $5.0$ \\
\cline{1-4}
standard model & $1$ & $1.0$ & $5.1$ \\
\hline
\hline
$SU(5)$ w/ ${\bf 5}$, ${\bf 24}$ & $-17$ & $0.74$ & $3.71$ & --- \\
\hline
$SU(5)$ w/ ${\bf 5}$, ${\bf 200}$ & $159$ & $2.3$ & $11.5$ & $42$ \\
\hline
$SU(5)$ w/ ${\bf 5}$, ${\bf 24}$, ${\bf 75}$ & $58$ & $1.6$ & $8.0$ & --- \\
\hline
$SU(5)$ w/ ${\bf 5}$, ${\bf 24}$, ${\bf 75}$, ${\bf 200}$ & $258$ & $2.8$ & $14.0$ & $14$ \\
\hline
$SO(10)$ w/ ${\bf 10}$, ${\bf 16}$, ${\bf 45}$ & $-35$ & $0.27$ & 1.34 & --- \\
\hline
$SO(10)$ w/ ${\bf 10}$, ${\bf 16}$, ${\bf 210}$ & $130$ & $2.1$ & $10.6$ & --- \\
\hline
$SO(10)$ w/ ${\bf 10}$, ${\bf 16}$, ${\bf 770}$ & $690$ & $4.4$ & $22.0$ & $3.9$ \\
\hline
\hline
SUSY-$SU(5)$ w/ ${\bf 5}$, $\overline{\bf 5}$, ${\bf 24}$ & $165$ & $2.3$ & $11.6$ & --- \\
\hline
SUSY-$SU(5)$ w/ ${\bf 5}$, $\overline{\bf 5}$, ${\bf 24}$, ${\bf 75}$ & $390$ & $3.4$ & $16.9$ & $3.6$ \\
\hline
SUSY-$SU(5)$ w/ ${\bf 5}$, $\overline{\bf 5}$, ${\bf 200}$ & $693$ & $4.4$ & $22.1$ & $0.85$ \\
\hline
SUSY-$SO(10)$ w/ ${\bf 10}$, ${\bf 16}$, $\overline{\bf 16}$, ${\bf 45}$, ${\bf 54}$ & $432$ & $3.5$ & $17.7$ & $11$ \\
\hline
SUSY-$SO(10)$ w/ ${\bf 10}$, ${\bf 16}$, $\overline{\bf 16}$, ${\bf 210}$ & $765$ & $4.6$ & $23.1$ & $1.8$ \\
\hline
SUSY-$SO(10)$ w/ ${\bf 10}$, ${\bf 16}$, $\overline{\bf 16}$, ${\bf 770}$ & $2445$ & $8.1$ & $40.7$ & $0.27$ \\
\hline
\hline
compactification scale $M_c$ (e.g.~\cite{Shafi}) & & \multicolumn{2}{|c||}{up to $\xi\sim100$} & \\
\hline
\end{tabular}
\end{center}
\caption{For models with various particle contents, the third column shows the effect $M_{Pl}\to M_{Pl}/\eta$ on the Planck scale entailed by the running (\ref{runningGN}) of Newton's constant. The fourth column gives numerical values for $\xi$ in (\ref{suppressionscale}) if the additional reduction factor $\sqrt{8\pi}$ from perturbative quantum gravity is taken into the suppression scale, as commonly done. Throughout we assume three generations of fermions, and grand unified models are characterized by their gauge group and Higgs content. In almost all cases -- and certainly in all models of our main interest, namely in non-supersymmetric grand unified models with several different Higgs multiplets at the unification scale -- reasonable values for $M_{Pl}$ are smaller than the naive value $1.2\times10^{19}\,\text{GeV}$, in some cases by as much as roughly an order of magnitude, i.e.~$1\leq\xi\lesssim O(10)$ in (\ref{suppressionscale}). The last column quantifies roughly how many orders of magnitude above the unification scale $M_X$ the grand unified gauge theories enter non-perturbative strong coupling regime $g_G(M_{np})=\sqrt{4\pi}$, a dash (---) indicating that the theory is asymptotically free; here, we have assumed $g_G(M_X)^2/4\pi=1/40$ for the non-supersymmetric and (conservatively) $g_G(M_X)^2/4\pi=1/30$ for the supersymmetric cases. When $M_{np}>M_{Pl}$, the unified field theory has a chance of describing Nature perturbatively up to the onset of quantum gravity at $M_{Pl}$.\label{etatable}}
\end{table}

Because they respect the symmetries of the theory and because quantum gravity effects \emph{are} mediated at the scale $M_{Pl}$, operators (\ref{dim5operators}) should be expected in the effective Lagrangian of grand unification and, although their sizes cannot be computed in this effective field theory point of view, they are expected a priori with natural Wilson coefficients of order $\vert c_i\vert\sim O(1)$ (for values much bigger than this, the proposed effective theory is not a good low-energy description and one loses perturbative control, whereas $\vert c_i\vert\ll1$ seemingly constitutes fine-tuning \cite{tHooft:1979bh}). And even though irrelevant for our purposes here, there are several known ways to generate these operators. For example, they arise from an ${\cal N}=1$ supergravity ultraviolet completion with non-canonical gauge kinetic function $f^{ab}(H_i)$ as the lowest order (non-trivial) terms in an expansion of $f^{ab}$ in $M_{Pl}$ \cite{Ellis:1985jn,Drees:1985bx} (in this scenario, supersymmetry needs to be broken in the grand unified theory at $M_{Pl}$ to conform to our non-supersymmetric analysis). Gravitational instantons can induce such effects also \cite{Perry:1978fd}. And in descending from higher-dimensional completions to four dimensions, spontaneous compactification generates operators (\ref{dim5operators}) suppressed by the compactification scale $M_c=M_{Pl}$ \cite{compactscale}. At any rate, lacking knowledge of quantum gravity, it seems most reasonable to assume the presence of effective operators (\ref{dim5operators}) with coefficients $c_i$ of order one.

\bigskip

We now describe the effect of the operators (\ref{dim5operators}) on the condition for gauge coupling unification. At the scale $M_X$ of grand unified symmetry breaking, some Higgs multiplets $H_i$ acquire nonzero vacuum expectation values (vevs). For simplicity and definiteness we assume that all Higgs fields, except for the multiplet containing the standard model Higgs, acquire vevs at the scale $M_X$; as will become clear, other multiplets, that get nonzero vevs only at lower scales, contribute proportionally less, but can in principle be treated equivalently (this assumption also avoids the introduction of further mass hierarchies into the model).

The vevs $\langle H_i\rangle$, acquired well above the electroweak scale, have to be invariant under the standard model subgroup $G_{321}=SU(3)_C \times SU(2)_L \times U(1)_Y\subset SU(5)$. For each multiplet $H_i$ that can occur in the dimension-5 operators (\ref{dim5operators}), this requirement determines its vevs $\langle H_i^{ab}\rangle$ up to an overall scale $v_i$ (see the Appendix). Replacing $H_i$ by their vevs, the operators (\ref{dim5operators}) modify the kinetic terms of the gauge bosons in the Lagrangian at the unification scale $M_X$ by adding to them:
\begin{eqnarray}
{\cal L}&=&-\frac{1}{4}G^a_{\mu\nu}G^{a\mu\nu}+\sum_i\frac{c_i}{4M_{Pl}}\,\langle H_i^{ab}\rangle\,G^a_{\mu\nu}G^{b\mu\nu}\nonumber\\
&=&-\frac{1}{4}(1+\epsilon_3)F^a_{\mu\nu}F^{a\mu\nu}_{SU(3)}-\frac{1}{4}(1+\epsilon_2)F^a_{\mu\nu}F^{a\mu\nu}_{SU(2)}-\frac{1}{4}(1+\epsilon_1)F_{\mu\nu}F^{\mu\nu}_{U(1)}~+~\ldots~,\label{noncanonicalterms}
\end{eqnarray}
where the ellipses denote the non-standard model gauge bosons of $SU(5)$, which become massive and (quasi) non-dynamical below $M_X$. The $\epsilon_s$ indicate the modifications to the gauge kinetic terms,
\begin{equation}\label{defepsilon}
\epsilon_s=\sum_i\frac{c_i}{M_{Pl}}v_i\delta^{(i)}_{s}~~~~(\text{for}~s=3,2,1)~;
\end{equation}
they generically differ for each factor $s=3,2,1$ of the standard model gauge group $G_{321}$ and depend on the Higgs content $H_i$ and on the sizes $c_iv_i/M_{Pl}$ of the vevs and Wilson coefficients relative to the suppression scale (thus, a Higgs that acquires its vev $v\sim M_I\ll M_X$ at an intermediate scale contributes much less than a Higgs at $M_X$). The $\delta_s^{(i)}$ are group theoretical factors (Clebsch-Gordan coefficients) specific to the embedding $G_{321}\subset SU(5)$ and characterize the possible standard model singlet vevs $\langle H_i\rangle$; they depend only on the representation $r_i$ of $H_i$ and are given in Table \ref{deltaSU5table}.

\begin{table}[htb]
\begin{center}
\begin{tabular}{c||c|c|c}
\hline
$SU(5)$ irrep $r$ & $\delta_1^{(r)}$ & $\delta_2^{(r)}$ & $\delta_3^{(r)}$ \\
\hline
\hline
${\bf 1}$ & $-1/\sqrt{24}$ & $-1/\sqrt{24}$ & $-1/\sqrt{24}$ \\
\hline
${\bf 24}$ & $1/\sqrt{63}$ & $3/\sqrt{63}$ & $-2/\sqrt{63}$ \\
\hline
${\bf 75}$ & $5/\sqrt{72}$ & $-3/\sqrt{72}$ & $-1/\sqrt{72}$ \\
\hline
${\bf 200}$ & $-10/\sqrt{168}$ & $-2/\sqrt{168}$ & $-1/\sqrt{168}$ \\
\hline
\end{tabular}
\end{center}
\caption{The Clebsch-Gordan coefficients $\delta_s$ in (\ref{defepsilon}) associated with the embedding $G_{321}\subset SU(5)$, for each irrep $r$ of $SU(5)$ that can occur as a multiplet $H_i$ in (\ref{dim5operators}). Sensible normalization conventions (see the Appendix), which ensure uniform treatment of different dimension-5 operators (\ref{dim5operators}), fix the values of all $\delta_s^{(r)}$ up to an overall sign for each $r$.\label{deltaSU5table}}
\end{table}

In the effective Lagrangian below the unification scale $M_X$, one would like to have canonically normalized gauge fields as opposed to (\ref{noncanonicalterms}), since it is the coupling constants associated with those that obey the familiar renormalization group (RG) equations ($\beta$--functions) and that are measured in low-energy experiments. This can be achieved by a finite (and usually small, see later) redefinition $A_{(s)}^{\mu}\to(1+\epsilon_s)^{1/2}A_{(s)}^{\mu}$ of the gauge fields associated with each standard model factor $s=3,2,1$, which has to be accompanied by a redefinition $g_s\to(1+\epsilon_s)^{-1/2}g_s$, so as not to affect the interaction strength. Gauge coupling unification requires that at the unification scale $M_X$ the couplings \emph{before} this redefinition meet at a common value $g_G=g_G(M_X)$, the gauge coupling of the unified group $G$ at the unification scale. In terms of the rescaled couplings $\alpha_s\equiv g_s^2/4\pi$, this reads:
\begin{equation}\label{unificationcondition}
(1+\epsilon_1)\alpha_1(M_X)=(1+\epsilon_2)\alpha_2(M_X)=(1+\epsilon_3)\alpha_3(M_X)~=~\frac{g_G^2}{4\pi}\equiv\alpha_G~.
\end{equation}
Under our assumption of one-step breaking at $M_X$ to the standard model, the running coupling functions $\alpha_s(\mu\leq M_X)$ are fixed by the fairly precise low-energy measurements \cite{Amsler:2008zzb} and their RG evolution \cite{Jones:1982}; the low-energy values and $\beta$--functions to one loop are given in the Appendix (\ref{runningcouplingsAPP})--(\ref{initialvaluesAPP}).

In Section \ref{sectionSU5results} we will be looking for non-supersymmetric models of $SU(5)$ grand unification (specified by their Higgs content $H_i$) which feature exact unification of the standard model gauge couplings $\alpha_s(\mu)$ under the unification condition (\ref{unificationcondition}). Such gauge coupling unification can happen naturally at scales $M_X>10^{16}\,\text{GeV}$, larger than normally expected, so as to escape the proton decay limit usually encountered in non-supersymmetric grand unification, and even at scales as large as the Planck scale. Through several criteria we will assess reasonableness of these models, in particular by checking whether their required Wilson coefficients have natural sizes $\vert c_i\vert\sim O(1)$ and also by looking at the modifications to the gauge kinetic terms in (\ref{noncanonicalterms}) (e.g., if $\epsilon_s<-1$ for any $s$, the gauge kinetic terms in the broken theory would have the wrong sign).

One other criterion for all models, necessary to sensibly claim unification, is the requirement that the masses of all non-standard model gauge bosons (``superheavy gauge bosons'') be close to the scale of grand unified symmetry breaking, i.e., close to the unification scale $M_X$. This is to ensure rightful use of the standard model RG equations with three differently evolving gauge couplings up to the scale $M_X$, where the gauge couplings are to unify at $g_G$ (\ref{unificationcondition}), and of the running of a single coupling of the unified gauge theory thenceforth with all gauge bosons being massless. At least this is necessary in the absence of intermediate scales, which is usually understood for $SU(5)$ unification models (see above). In $SU(5)$, all superheavy gauge bosons get equal masses from grand unified symmetry breaking,
\begin{equation}\label{gaugebosonmassSU5}
M_\text{gb}=g_G\,\sqrt{\sum_i\frac{C_2(r_i)}{12}v_i^2}~,
\end{equation}
where the sum now runs over all Higgs multiplets $i$ in the theory that acquire nonzero vev $v_i$ at $M_X$, with the Casimir invariants $C_2$ of the representations $r_i$ (cf.~Table \ref{C2table}). For definiteness, and to conform to standard treatment, we take the above requirement then to be
\begin{equation}\label{gbmassrequirement}
M_\text{gb}~=~M_X~.
\end{equation}
Furthermore, since Higgses in any of the representations ${\bf 24}$, ${\bf 75}$, ${\bf 200}$ can achieve grand unified symmetry breaking $SU(5)\to G_{321}$, we will make the simplifying assumption that the theory only contain Higgses able to form the dimension-5 operators (\ref{dim5operators}); the more general case is not harder to treat but increases the particle content of the models and requires larger Wilson coefficients to achieve the desired effects.

A requirement similar to (\ref{gbmassrequirement}) should be put on the masses of the superheavy scalars as well, i.e.~the Higgs fields that acquire nonzero vev at $M_X$, which we, however, will neglect since generically most superheavy scalars are more massive than the (lightest of the) superheavy gauge bosons and since the superheavy scalars do not influence the running of the standard model gauge couplings at one loop. Also, we neglect heavy particle thresholds \cite{Weinberg:1980wa,Lavoura:1993su}, to demonstrate unification due specifically to the dimension-5 operators (\ref{dim5operators}) as a proof of principle, cleanly separated from other effects.

\bigskip

In this paper we focus on non-supersymmetric grand unification. Contrary to the standard lore, and this is one of our main findings (Sections \ref{sectionSU5results} and \ref{protondecaysection}), the altered unification condition (\ref{unificationcondition}) can yield successful non-supersymmetric unification models which satisfy the proton lifetime constraint. Furthermore, non-supersymmetric grand unified models hold the attractive hope of describing physics up to the Planck scale in a perturbative way, whereas supersymmetric models commonly become strongly coupled before reaching the Planck scale. For several non-supersymmetric as well as supersymmetric grand unified models, the right column in Table \ref{etatable} indicates roughly how many orders of magnitude above the unification scale $M_X$ the theories become non-perturbative (i.e., near a Landau pole) and illustrates that generally only non-supersymmetric models are safe in that regard, requiring roughly 3 orders of magnitude between $M_X$ and the non-perturbative regime $M_{np}$. Furthermore, as they contain fewer scalars and fermions, the running (\ref{runningGN}) of Newton's constant and the associated change of the Planck scale $M_{Pl}$ from its naive or reduced value tend to be smaller in the non-supersymmetric models (see Table \ref{etatable}), diminishing uncertainties in these models \cite{ourprl}.

\section{Models and unification results in the non-supersymmetric ${\bf{SU(5)}}$ case}\label{sectionSU5results}
In this section, we will quantitatively examine the effects of multiple gravitationally induced dimension-5 operators (\ref{dim5operators}) and of the subsequently modified condition (\ref{unificationcondition}) for gauge coupling unification in several non-supersymmetric $SU(5)$ models. All models are presumed to have the minimal fermionic content of three standard model families, but differ in their Higgs content responsible for grand unified symmetry breaking; without loss (see above) we only consider Higgs representations ${\bf 1}$, ${\bf 24}$, ${\bf 75}$, ${\bf 200}$. As the Higgs multiplets under consideration are suitable for breaking $SU(5)$ down to the standard model, we do not consider the Higgs (scalar) potential in these unified theories explicitly; rather, we take as the parameters of the models the Higgs vevs $v_i$ directly, which are acquired in grand unified symmetry breaking as a consequence of the Higgs potential, and the Wilson coefficients $c_i$ of the dimension-5 operators (\ref{dim5operators}).

Our method is then as follows. After specifying a grand unified model by its Higgs content, we scan its parameter space $\left\{c_i,v_i\right\}$ for points that, below the breaking scale $M_X$, yield the actual running gauge couplings $\alpha_s(\mu)$ of the standard model. Phrased in a bottom-up language, we are looking for points $\left\{c_i,v_i\right\}$ in parameter space that result in a unification condition (\ref{unificationcondition}), according to which the actually observed gauge couplings $\alpha_s(\mu)$ of the standard model unify, while simultaneously requirement (\ref{gbmassrequirement}) holds, namely that the superheavy gauge boson masses (\ref{gaugebosonmassSU5}) be equal to the unification scale. In fact, any such point $\left\{c_i,v_i\right\}$ determines $M_X$ and the unified gauge coupling $g_G\equiv g_G(M_X)$ uniquely; see the Appendix below (\ref{howtosolve}) where we also outline how all points $\left\{c_i,v_i\right\}$ may be found.

In the next subsection we look at the known scenario in which the theory contains only one Higgs that can form a dimension-5 operator, as a warm-up and comparison to established results \cite{Hill:1983xh,Shafi,Chakrabortty:2008zk}. For the case of two or more operators (\ref{dim5operators}), we find points in parameter space that yield physically viable unification. These effects are described in detail exemplarily for scenarios with two dimension-5 operators in Section \ref{twoHiggsSU5subsection}, before moving on to a more general treatment of models with any number of dimension-5 operators where we also pay special attention to naturalness.

\subsection{Unification in models with one dimension-5 operator}\label{SU5oneoperatorsection}
When the modifications $\epsilon_s$ to the gauge kinetic terms (\ref{noncanonicalterms}) come from only one Higgs (\ref{dim5operators}), their ratio $\epsilon_1:\epsilon_2:\epsilon_3$ is completely determined, see (\ref{defepsilon}). For the given running gauge coupling functions $\alpha_s(\mu)$ of the standard model, this ratio uniquely determines, via the unification condition (\ref{unificationcondition}), the possible unification scale $M_X$, the unified coupling $\alpha_G=g_G^2/4\pi$ and the required absolute sizes of all $\epsilon_s$; subsequently, the required Higgs vev $v$ and Wilson coefficient $c$ necessary for unification may be computed. It then has to be checked whether all these values are physically reasonable.

For numerical results, see Table \ref{results1opSU5table} which shows these quantities for the $SU(5)$ models with a sole Higgs multiplet ${\bf 24}$, ${\bf 75}$, or ${\bf 200}$. A sole singlet ${\bf 1}$ cannot modify condition (\ref{unificationcondition}) to make the standard model gauge couplings unify since $\epsilon_1=\epsilon_2=\epsilon_3$ in this case, neither can a singlet vev break the unified symmetry in the first place. Although the required modifications $\epsilon_s$ to the gauge kinetic terms are not problematic (in particular, $\epsilon_s>-1$) and even small, the necessary Wilson coefficients $c$ are big in the case of a ${\bf 75}$ and especially large for a ${\bf 24}$ Higgs, regardless of any reasonable choice of suppression scale; operators (\ref{dim5operators}) with such large coefficients are not expected naturally in an effective theory and might preclude a perturbative treatment. Furthermore, in the case of a ${\bf 24}$ the unification scale $M_X$ would be far too low to satisfy current constraints on the proton lifetime, and possibly slightly too low for a ${\bf 75}$ (for details, see Section \ref{protondecaysection}). For a ${\bf 200}$ Higgs the Wilson coefficient $c$ has natural size and, depending on convention (\ref{suppressionscale}), the unification scale may still lie below or at the Planck scale $M_{Pl}$, which may be taken as a hint to a scenario of simultaneous gauge-gravity unification (see Section \ref{gaugegravitysection}).

\begin{table}[htb]
\begin{center}
\begin{tabular}{c||c|c|c|c|c||c|c||c|c|c|c||c}
\hline
$H$ irrep & $M_X/\text{GeV}$ & $1/\alpha_G$ & $c$ & $v/\text{GeV}$ & $\text{max}_s\,\vert\epsilon_s\vert$ & $M_X^{(2\text{lp})}/\text{GeV}$ & $c^{(2\text{lp})}$ & $\xi=1$ & $\xi_\text{red}=\sqrt{8\pi}$ & $\xi^\text{run}$ & $\xi^\text{run}_\text{red}$ & $O$ in (\ref{estimatewithnumerics}) \\
\hline
\hline
${\bf 1}$ & \multicolumn{7}{c||}{unification by mechanism (\ref{unificationcondition}) (and symmetry breaking) impossible} & \multicolumn{4}{c||}{}&\\
\hline
${\bf 24}$ & $4.6\times10^{13}$ & $40.6$ & $18700/\xi$ & $1.3\times10^{14}$ & $0.076$ & $4.0\times10^{13}$ & $19200/\xi$ & $1$ & $5$ & $0.74$ & $3.71$ & $0.072$ \\
\hline
${\bf 75}$ & $8.1\times10^{15}$ & $43.3$ & $-129/\xi$ & $1.8\times10^{16}$ & $0.116$ & $3.7\times10^{15}$ & $-248/\xi$ & $1$ & $5$ & $1.4$ & $6.9$ & $0.087$ \\
\hline
${\bf 200}$ & $5.2\times10^{18}$ & $53.4$ & $0.53/\xi$ & $1.1\times10^{19}$ & $0.363$ & $9.8\times10^{17}$ & $2.6/\xi$ & $1$ & $5$ & $2.3$ & $11.5$ & $0.23$ \\
\hline
\end{tabular}
\end{center}
\caption{For all suitable Higgs irreps, this table shows the parameters a non-supersymmetric $SU(5)$ model must have if unification of the standard model happens by means of mechanism (\ref{unificationcondition}) with only one dimension-5 operator (\ref{dim5operators}). Columns 2--6 are obtained by using the one-loop $\beta$--functions for the standard model gauge couplings, columns 7--8 by two-loop for comparison. The size of the required Wilson coefficients $c$ depends on the chosen suppression scale (\ref{suppressionscale}) parametrized by $\xi$, four plausible values of which are given for each model: naive or reduced Planck scale, either or not taking the running of Newton's constant into account, cf.~Table \ref{etatable}. For the last column, see below estimate (\ref{estimatewithnumerics}).\label{results1opSU5table}}
\end{table}

These unification results largely agree with \cite{Chakrabortty:2008zk}. There is, however, some disagreement on a sensible normalization among the dimension-5 operators with Higgses in different representations, which directly affects the required sizes of the Wilson coefficients $c$. We have chosen to write all possible dimension-5 operators in a common form (\ref{dim5operators}) with uniform normalization conventions (see the Appendix), and believe that this allows for a sensible cross-comparison between the Wilson coefficients of different operators. Furthermore, within these conventions, $\vert c\vert\sim1$ is believed to be a natural size for the dimension-5 operators from an effective field theory point of view (see Section \ref{sectionSU5setup} and the Appendix). Since the low-energy gauge couplings $\alpha_s(m_Z)$ had not been measured well at the time and were therefore partly treated as free parameters, Table \ref{results1opSU5table} cannot be directly compared to the pioneering work \cite{Hill:1983xh,Shafi}, that looked only at the case of a sole ${\bf 24}$ Higgs, but the results tend in the same direction.

Table \ref{results1opSU5table} also shows results obtained by using two-loop renormalization group equations \cite{Jones:1982} (neglecting matter fields) instead of (\ref{runningcouplingsAPP}). A full two-loop treatment would further incorporate the effect of thresholds, as partly done in \cite{Chakrabortty:2008zk}, and couplings to the fermion and Higgs fields, increasing arbitrariness and decreasing predictivity \cite{Lavoura:1993su,Weinberg:1980wa}; there would also be some renormalization scheme dependence. The one-loop values are not significantly altered, so a one-loop analysis can show, with reasonable numerical accuracy, whether the modified condition (\ref{unificationcondition}) can make the standard model gauge couplings unify, as a proof of principle.

Furthermore, endowing the measured low-energy couplings (\ref{initialvaluesAPP}) with error bars would put uncertainties on the $M_X$ and $c$ necessary to achieve unification under these initial conditions; for example, in the case of a {\bf 75}, the error bars (less than 4\% \cite{Amsler:2008zzb}) given in (\ref{initialvaluesAPP}) result in $M_X=(8.1\pm6.9)\times10^{15}\,\text{GeV}$ and $c=(-129\pm93)/\xi$, which would not significantly influence whether (natural) unification is considered a reasonable possibility or not. (In a previous paper \cite{ourprl}, however, the relative effect from the uncertainty on the necessary Wilson coefficients $c$ was seen to be much bigger, as very small Wilson coefficients are required in supersymmetric unification models in the first place.) Besides being predictive, our analysis shows the effect due solely to the modified unification condition (\ref{unificationcondition}) cleanly separated from other effects like, e.g., threshold corrections \cite{Lavoura:1993su,Weinberg:1980wa}.

\subsection{Unification in models with two dimension-5 operators}\label{twoHiggsSU5subsection}
In the case of multiple dimension-5 operators (\ref{dim5operators}) the modification to the gauge kinetic terms (\ref{noncanonicalterms}) is a linear combination of the effects from single operators, weighted by the Wilson coefficients and Higgs vevs, see (\ref{defepsilon}). The idea is that now, as one can vary these contributions with the model parameters in a continuous way, one can find subsets of parameter space that result in perfect gauge coupling unification at continuously variable scales $M_X$.

In the case of two dimension-5 operators (not in the same representation) these subsets are two-dimensional, by counting the number of parameters ($c_i$ and $v_i$ for each Higgs $i$) less the two unification constraints (\ref{unificationcondition}): $2\cdot2-2=2$. To emphasize the possibility of varying $M_X$ continuously, we choose to parametrize these subsets by the unification scale $M_X$ itself and by the ratio $v_1:v_2$ of the two Higgs vevs, which, without fine-tuning the Higgs potential, should probably be within an order of magnitude of each other.

\begin{figure}
\includegraphics[scale=0.8]{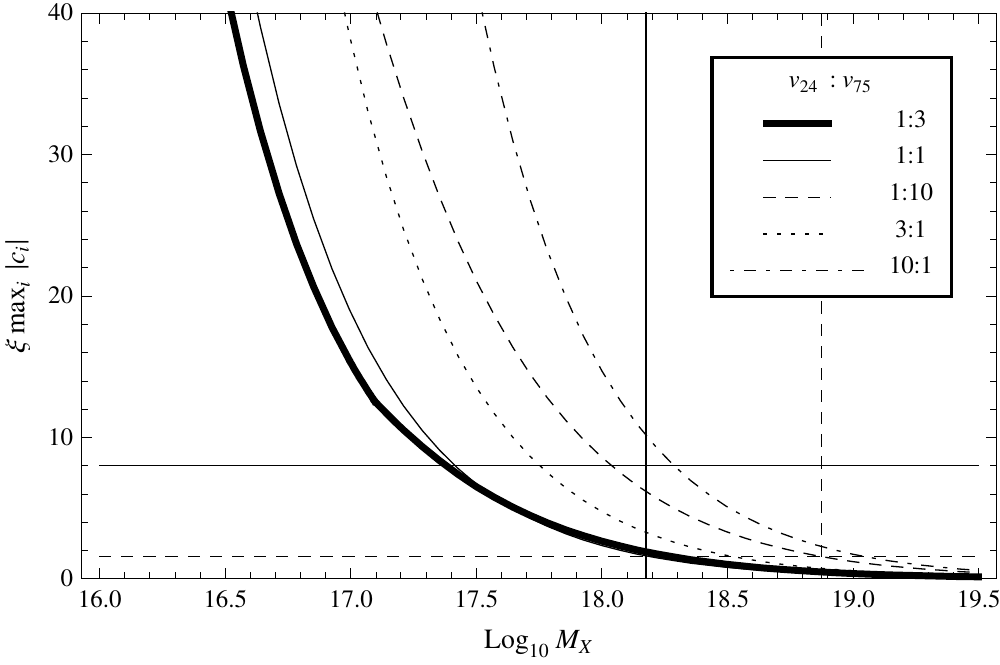}~~~~~~~~~~\includegraphics[scale=0.8]{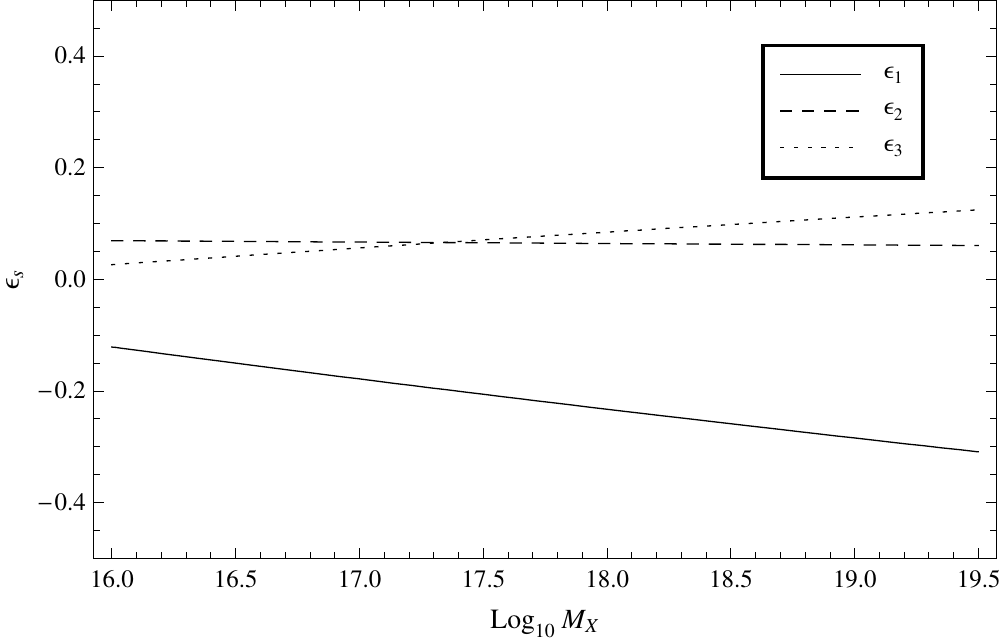}
\caption{Numerical gauge coupling unification results for the $SU(5)$ model with a ${\bf 24}$ and a ${\bf 75}$ Higgs. Left panel: For different fixed ratios $v_{24}:v_{75}$ of the Higgs vevs, the curves characterize the size of the Wilson coefficients $c_i$ necessary to achieve unification at $M_X$. The straight horizontal dashed and solid lines indicate two possible choices for the parameter $\xi^\text{run}=1.6$ and $\xi^\text{run}_\text{red}=8.0$ in (\ref{suppressionscale}) (cf.~also Table \ref{etatable}), and the vertical lines the corresponding suppression scales $M_{Pl}=1.2\times10^{19}\,\text{GeV}/\xi$ (it may not make sense to consider scales $M_X$ larger than $M_{Pl}$). Right: The $\epsilon_s$ in (\ref{unificationcondition}) necessary to achieve unification at a given $M_X$ (not dependent on the ratio $v_{24}:v_{75}$ of Higgs vevs).\label{SU5w24and75-cmax-epsilons}}
\end{figure}

Numerical results for the model with a ${\bf 24}$ and a ${\bf 75}$ Higgs are shown in Fig.~\ref{SU5w24and75-cmax-epsilons}. For a given choice of maximally (naturally) acceptable Wilson coefficients $\max\{\vert c_{24}\vert,\vert c_{75}\vert\}$, the lowest unification scale $M_X$ can be achieved for the (not finely-tuned) vev ratio of roughly $v_{24}:v_{75}=1:3$ (solid bold curve in the figure). For example, demanding $\vert c_{24}\vert,\vert c_{75}\vert<1$ and choosing the suppression scale $M_{Pl}=1.2\times10^{19}\,\text{GeV}/\xi_\text{red}=2.4\times10^{18}\,\text{GeV}$, then any unification scale $M_X\geq5\times10^{17}\,\text{GeV}$ can be achieved, whereas, if one allows $\vert c_{24}\vert,\vert c_{75}\vert<5$ with $M_{Pl}=1.2\times10^{19}\,\text{GeV}/\xi^\text{run}_\text{red}=1.5\times10^{18}\,\text{GeV}$, then any $M_X\geq3\times10^{16}\,\text{GeV}$ is possible, getting close to the contraint on proton decay (Section \ref{protondecaysection}). The right panel shows the modifications to the gauge kinetic terms (\ref{noncanonicalterms}) necessary to achieve unification at $M_X$ in this model; in no case are big $\epsilon_s$ required that would invalidate the analysis (in particular, $\epsilon_s>-1$ always).

\begin{figure}
\includegraphics[scale=0.8]{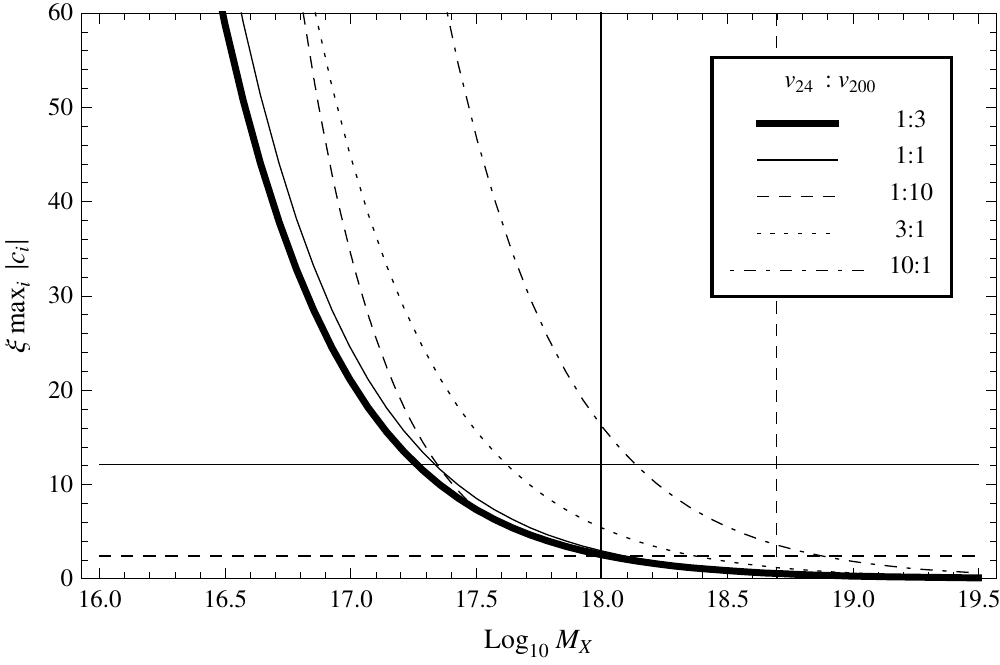}~~~~~~~~~~\includegraphics[scale=0.8]{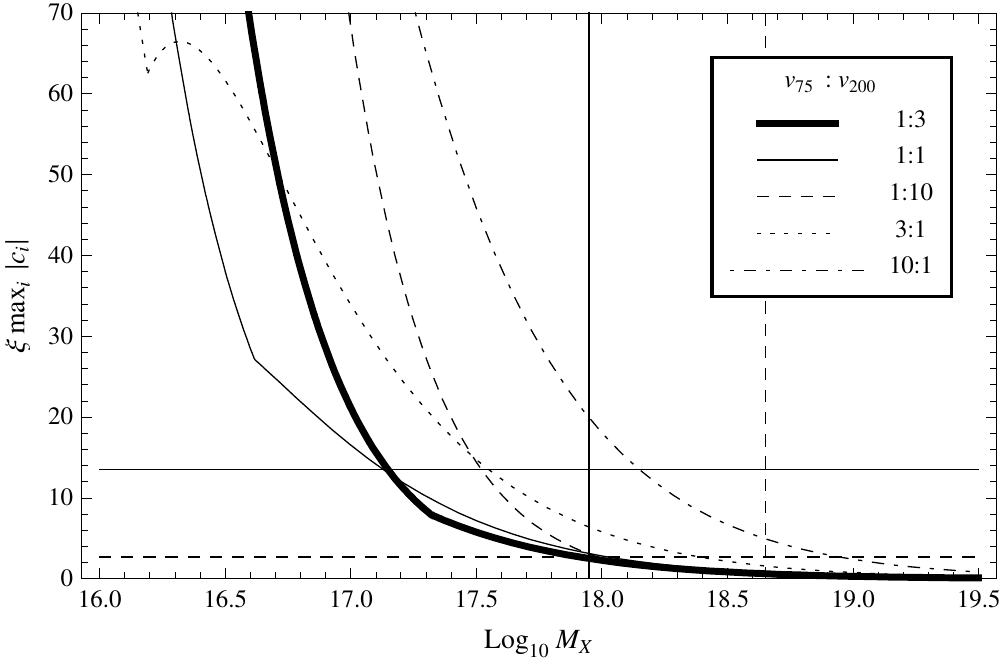}
\caption{Similar to the left panel of Fig.~\ref{SU5w24and75-cmax-epsilons}, but for the models with ${\bf 24}$, ${\bf 200}$ Higgses (left; $\xi^\text{run}=2.42$, $\xi^\text{run}_\text{red}=12.1$) and ${\bf 75}$, ${\bf 200}$ Higgses (right; $\xi^\text{run}=2.68$, $\xi^\text{run}_\text{red}=13.6$). For the ${\bf 24}$, ${\bf 200}$ model, the ratio of Higgs vevs that achieves any chosen $M_X$ with the smallest Wilson coefficients is roughly $v_{24}:v_{200}=1:3$ (solid bold curve), and for the ${\bf 75}$, ${\bf 200}$ model this ratio depends on $M_X$ and lies between $1:3$ and $3:1$, requiring no fine-tuning of the Higgs potential. In both models, $\vert\epsilon_s\vert<0.5$ at all unification scales $M_X$ shown here (cf.~right panel of Fig.~\ref{SU5w24and75-cmax-epsilons}).\label{SU5w24and200_w75and200}}
\end{figure}

Similar results obtain for the two models with ${\bf 24}$, ${\bf 200}$ resp.~${\bf 75}$, ${\bf 200}$ Higgses, see Fig.~\ref{SU5w24and200_w75and200}. Within the allowances of the previous paragraph, the widest unification scales achievable are $M_X\geq1.4\times10^{16}\,\text{GeV}$, namely for the model with ${\bf 75}$, ${\bf 200}$ Higgses, when accepting $\max\{\vert c_{75}\vert,\vert c_{200}\vert\}<5$ and assuming $M_{Pl}=1.2\times10^{19}\,\text{GeV}/\xi^\text{run}_\text{red}$. Requiring $\max_i\vert c_i\vert<1$ and assuming the non-reduced Planck scale, one can achieve unification at any $M_X\geq8\times10^{17}\,\text{GeV}$ (roughly the same for both models).

In models with two dimension-5 operators (\ref{dim5operators}) involving Higgs multiplets in \emph{identical} irreps (e.g.~two ${\bf 24}$ Higgses), one cannot shift the unification scale continuously since the ratio between the $\epsilon_s$ cannot be varied continuously. In fact, only the unification scales $M_X$ given in Table \ref{results1opSU5table} are possible, with Higgs vevs and Wilson coefficients modified by $O(1)$ factors. Furthermore, in models with one singlet and one non-singlet Higgs (e.g., one ${\bf1}$ and one ${\bf24}$ Higgs) the unification equations do not have physically sensible solutions (formally, the equations yield $\alpha_s(M_X)\approx0$ at any $M_X$).

\subsection{Effects in general models}\label{generalmodelssection}
The behavior from Figs.~\ref{SU5w24and75-cmax-epsilons} and \ref{SU5w24and200_w75and200} is generic for models with at least two dimension-5 operators, as we will see now. In particular, there are no points in parameter space which yield unification at some given $M_X$ for significantly smaller Wilson coefficients than shown in these figures. This is true in models with at least two dimension-5 operators (\ref{dim5operators}) and involving Higgses in at least two unequal irreps from the set ${\bf 24}$, ${\bf 75}$, ${\bf 200}$; in all other models, acceptable exact unification via mechanism (\ref{unificationcondition}) is impossible for general $M_X$, as we have just seen, and can at most happen at the discrete scales $M_X$ from Table \ref{results1opSU5table}.

Writing $c\equiv\sum_{i=1}^h c_i$ and $v\equiv\sum_{i=1}^h v_i$ with $h$ the number of Higgs multiplets at the grand unification scale, equations (\ref{unificationcondition}), (\ref{defepsilon}), (\ref{gbmassrequirement}) and (\ref{gaugebosonmassSU5}) together yield for the Wilson coefficients $c_i$:
\begin{eqnarray}\label{wilsoncoefficientansatz}
c_i=\frac{c_i}{c}\frac{M_{Pl}}{M_X}\sqrt{4\pi\alpha_G(M_X)}\sqrt{\sum_{j=1}^h\frac{C_2(r_j)}{12}\left(\frac{v_j}{v}\right)^2}\left(\frac{\alpha_G(M_X)}{\alpha_s(M_X)}-1\right)\left(\sum_{k=1}^{h}\frac{c_k}{c}\frac{v_k}{v}\delta_s^{(k)}\right)^{-1}~~~~(\text{for any}~s=1,2,3)~.
\end{eqnarray}
(Higgs fields in representations $r_i\notin\{{\bf1},{\bf24},{\bf75},{\bf200}\}$ can easily be accommodated by setting $c_i\equiv\delta_s^{(i)}\equiv0$). Then setting $v_i/v\approx1/h$ (no large hierarchies between vevs $v_i$), $\max_i\vert c_i/c\vert\gtrsim O(1-5)/h\geq 1/h$ (the constant of proportionality depends on the hierarchy between and the signs of the $c_i$), $\alpha_G(M_X)\gtrsim1/50$ (typical for non-supersymmetric unification; see also Table \ref{results1opSU5table}) and $\sum_j C_2(r_j)=h\overline{C}_2\gtrsim5h$ (see Table \ref{C2table}), and using that (\ref{wilsoncoefficientansatz}) is valid for all $s=1,2,3$ gives
\begin{eqnarray}
\max_i\vert c_i\vert&\gtrsim&\frac{1}{h}\frac{M_{Pl}}{M_X}\sqrt{\frac{4\pi}{50}}\sqrt{\frac{5h}{12}\frac{1}{h^2}}~\left|\frac{\alpha_G(M_X)}{\alpha_s(M_X)}-1\right|~\left(\max_{s'}\left|\sum_k\frac{c_k}{c}\frac{1}{h}\delta_{s'}^{(k)}\right|\right)^{-1}\nonumber\\
&\gtrsim&\frac{1}{\sqrt{3h}}\frac{M_{Pl}}{M_X}~\left(\frac{1}{3}\sum_{s=1}^3\left|\frac{\alpha_G(M_X)}{\alpha_s(M_X)}-1\right|\right)~\left(\max_{s',k}\left|\delta_{s'}^{(k)}\right|\right)^{-1}~.
\end{eqnarray}
The last factor depends on the group theory constants $\delta_s^{(r)}$ associated with the embedding of the standard model into $SU(5)$ and numerically equals $(10/\sqrt{168})^{-1}$ (see Table \ref{deltaSU5table}); the other factor largely characterizes how well (or how badly) the actual running couplings $\alpha_s(\mu)$ of the standard model unify without any modification to the unification condition since it can be estimated with (\ref{unificationcondition}):
\begin{equation}
\sum_{s=1}^3\left|\frac{\alpha_G(M_X)}{\alpha_s(M_X)}-1\right|\approx2\left|\frac{\alpha_1(M_X)-\alpha_2(M_X)}{\alpha_1(M_X)+\alpha_2(M_X)}\right|+\left\{\begin{matrix}\alpha_1\to\alpha_2\\\alpha_2\to\alpha_3\end{matrix}\right\}+\left\{\begin{matrix}\alpha_1\to\alpha_3\\\alpha_2\to\alpha_1\end{matrix}\right\}\gtrsim0.25\,~~~(\text{for}~10^{15}\leq M_X/\text{GeV}\leq10^{19})~,
\end{equation}
yielding finally:
\begin{equation}\label{estimatemaxci}
\max_i\vert c_i\vert~\gtrsim~\frac{1}{15\sqrt{h}}\frac{M_{Pl}}{M_X}~~~~~(\text{for}~10^{15}\leq M_X/\text{GeV}\leq10^{19})~.
\end{equation}

This estimate captures unification according to the modified unification condition (\ref{unificationcondition}) pretty well for models with several Higgs multiplets (as specified above), as can be seen from numerical studies: from the numerical behavior, to achieve equality in (\ref{estimatemaxci}), $1/15$ should be replaced by some factor of order $O(0.08\to4)$, depending on the Higgs content of the model, on the ratio of vevs and $c_i$'s, and on $M_X$. The following lower bound can be strict in some models, but has leeway in most situations:
\begin{equation}\label{estimatewithnumerics}
\max_i\vert c_i\vert~\gtrsim~\frac{O(0.1)}{\sqrt{h}}\frac{M_{Pl}}{M_X}~=~\frac{O(0.1)}{\xi\sqrt{h}}\,\frac{1.2\times10^{19}\,\text{GeV}}{M_X}~~~~~(\text{for all}~10^{13}\leq M_X/\text{GeV}\leq10^{20})~.
\end{equation}
Furthermore, unification at any given $M_X$ \emph{can} be achieved in any such model when allowing coefficients $c_i$ of the size (\ref{estimatewithnumerics}) with the constant being $O(0.5)$.

\begin{figure}
\includegraphics[scale=0.8]{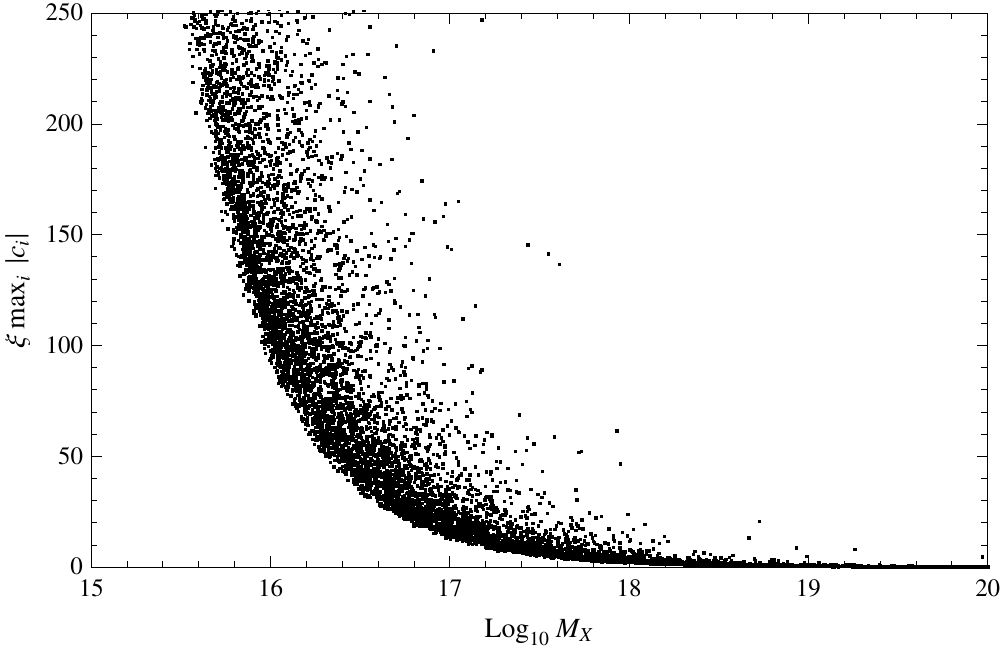}~~~~~~~~~~\includegraphics[scale=0.8]{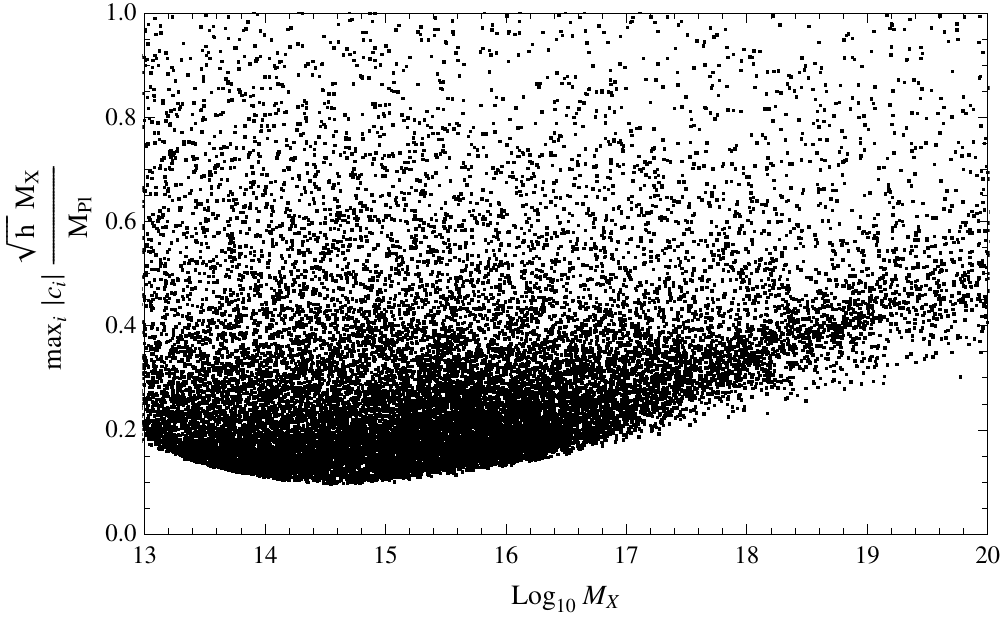}
\caption{For a model with the three ($h=3$) Higgs multiplets ${\bf24}$, ${\bf75}$ and ${\bf200}$, a random sample is taken of $30,000$ points $\{c_i,v_i\}$ of the parameter space which yield gauge coupling unification. The point plot on the left characterizes the required sizes of the Wilson coefficients $c_i$ to achieve unification, as in Figs.~\ref{SU5w24and75-cmax-epsilons} and \ref{SU5w24and200_w75and200}; unification at smaller scales $M_X$ requires proportionally larger Wilson coefficients. To illustrate the validity of (\ref{estimatewithnumerics}), the right panel shows, for each sample point, the $O(0.1)$ constant to achieve equality in (\ref{estimatewithnumerics}); numerical gauge coupling unification at any $10^{13}\,\text{GeV}<M_X<10^{20}\,\text{GeV}$ can be achieved by choosing $0.5$ as this constant, and a lower bound is obtained by the choice $0.1$. However, this should not be mistaken: unification at small $M_X\lesssim10^{16}\,\text{GeV}$ requires large dimensionless Wilson coefficients $c_i$ (see left panel).\label{general3estimatefigure}}
\end{figure}

For justification of (\ref{estimatewithnumerics}) and its $O(0.1)$ factor in a model with three Higgs multiplets, see Fig.~\ref{general3estimatefigure}, especially the right panel where the $O(0.1)$ factor to achieve equality in (\ref{estimatewithnumerics}) has been plotted for $30,000$ randomly sampled points of the parameter space that yield unification. (One can formally achieve small Wilson coefficients by making the number $h$ of Higgs multiplets large, but these models face issues such as Landau poles close to the unification scale and small Higgs vevs which lead to scalar masses potentially far below the unification scale.) An estimate like (\ref{estimatewithnumerics}) allows one to quickly judge whether, for some given Higgs content, unification via (\ref{unificationcondition}) at a desired scale $M_X$ is possible with natural-size Wilson coefficients $c_i$. Models containing Higgs multiplets in representations $r\notin\{{\bf24},{\bf75},{\bf200}\}$ have larger required $c_i$ than models without such multiplets, see (\ref{wilsoncoefficientansatz}). Estimate (\ref{estimatewithnumerics}) is also quite accurate at the unification point $M_X$ in models with only one dimension-5 operator, see last column of Table \ref{results1opSU5table}; however, in those models, $M_X$ cannot be continuously shifted to other values as in the models considered in this subsection here.

The ingredients of the unification mechanism discussed in this paper are well illustrated by (\ref{estimatewithnumerics}): For models with Higgs multiplets in at least two of the irreps ${\bf24}$, ${\bf75}$, ${\bf200}$, unification at continuously variable scale $M_X$ may be achieved due to the presence of gravitationally induced effective dimension-5 operators (\ref{dim5operators}) in the grand unified theory, since two or more such operators allow for a continuous set of solutions to unification condition (\ref{unificationcondition}). To actually achieve unification at a given $M_X$, however, these operators need to have the right sizes $c_i$, see (\ref{estimatewithnumerics}) for an estimate, and these sizes influence whether one considers unification at $M_X$ possible in a natural way (requiring, very roughly, $\vert c_i\vert\sim O(0.1\to10)$ or so). The sizes of the $c_i$ are further directly affected by the choice of the Planck scale $M_{Pl}$ (\ref{suppressionscale}) that suppresses the dimension-5 operators (\ref{dim5operators}). Reasonable choices in (\ref{estimatewithnumerics}) are $\xi=1$ (naive Planck scale) or, more commonly, $\xi_\text{red}=\sqrt{8\pi}\approx5$ (reduced Planck scale), possibly further enhanced by factors $O(1.5\to5)$ if the running of Newton's constant (\ref{runningGN}) in such models is taken into account as another, additional gravitational effect (cf.~$\eta$ in Table \ref{etatable}). These latter considerations about the appropriate choice of the Planck scale merely influence the required numerical values of the Wilson coefficients $c_i$ by factors of order $O(1\to25)$, and thus are only secondary to the possibility of continuous variability of the unification condition (\ref{unificationcondition}), which facilitates exact gauge coupling unification for some values of the $c_i$ in the first place.

In this section we have discussed how models can achieve exact gauge coupling unification through unification condition (\ref{unificationcondition}) and whether this is possible with natural coefficients. The former is a formal numerical requirement which grand unified theories ought to obey. But it is a further question whether those models are physically sensible and viable, e.g.~if they are allowed by experimental constraints. In the following two Sections \ref{protondecaysection} and \ref{gaugegravitysection}, we will look at some of the physics of these unification models.

\section{Non-supersymmetric unification near the proton lifetime limit}\label{protondecaysection}
In this and in the following section we will demonstrate and analyze two possible implications of models which employ the unification mechanism just described.

Here, we display and examine models which allow non-supersymmetric unification of the standard model into $SU(5)$ without intermediate symmetry breaking scales, while easily avoiding constraints from proton decay, contrary to a widely held belief. This scenario seems attractive as it does not require presently unobserved supersymmetry and as non-supersymmetric models offer the possibility of describing physics up to the onset of quantum gravity in a perturbative way, see the right column in Table \ref{etatable} and the last paragraph of Section \ref{sectionSU5setup}. However, the proposed models are not completely minimal in their Higgs content as mechanism (\ref{unificationcondition}) requires at least two multiplets for continuously shifting $M_X$ around (the minimal choice being a ${\bf24}$ and a ${\bf75}$ Higgs), whereas one ${\bf24}$ alone would already be sufficient to break the grand unified symmetry; nevertheless, as described in Section \ref{sectionSU5setup}, when this Higgs content is assumed, the presence of the gravitational operators (\ref{dim5operators}) is to be expected and does not further detract from minimality.

In non-supersymmetric models, proton decay is mediated by gauge $d=6$ operators (baryon number violating operators after integrating out the superheavy gauge fields) and Higgs $d=6$ operators (after integrating out superheavy Higgses). The latter are strongly dependent on the Higgs sector and Yukawa couplings; since they are, apart from the potentially dangerous triplet route, generically less important than the gauge $d=6$ operators (see \cite{protondecayreviewNath2006} for a review), we will concentrate on the gauge contribution for the following estimates (also neglecting potential ``textures'' in flavor space that could partially rotate away the effective gauge $d=6$ interactions, yielding a slower decay rate and weaker bounds \cite{HowLongCouldWeLive}). Under these assumptions, the proton decay rate $1/\tau_{(p\to e^+\pi^0)}$ through the dominant (in our models) channel $p\to e^+\pi^0$ is related to the superheavy gauge boson masses $M_X$ (\ref{gbmassrequirement}) by
\begin{equation}\label{protondecaytime}
\tau_{(p\to e^+\pi^0)}~=~O(1)\,\frac{M_X^4}{\alpha_G^2\,m_p^5}~,
\end{equation}
with the proton mass $m_p$ and we take the $O(1)$ constant of proportionality to be $1$ for the following. The current experimental bound on this decay channel is \cite{Amsler:2008zzb,protondecay92ktonyearbound}
\begin{equation}\label{decayratebound}
\tau_{(p\to e^+\pi^0)}~>~f\cdot10^{33}\,\text{years}~~~\text{with}~~f=1.6~~\text{or}~~f=5.4~.
\end{equation}
This sets a proton lifetime bound on the unification scale $M_X$ in non-supersymmetric models:
\begin{equation}\label{MXprotonconstraint}
M_X~>~\left(40\alpha_G\right)^{1/2}\,\left(\frac{f}{1.6}\right)^{1/4}\,2.4\times10^{15}\,\text{GeV}~\approx~(1.9-3.8)\times10^{15}\,\text{GeV}~,
\end{equation}
where the latter illustrates the range given by (\ref{decayratebound}) and $\alpha_G=1/60-1/30$ (typical for non-supersymmetric models using (\ref{unificationcondition}), cf.~also Table \ref{results1opSU5table}). Within the next ten years, if proton decay remains unobserved, improvements in the bound (\ref{decayratebound}) up to $f=100$ are expected \cite{protondecay10yearprojection}, constraining $M_X>8\times10^{15}\,\text{GeV}$.

\bigskip

One can now see that all of our models that achieve natural gauge coupling unification, i.e.~with Wilson coefficients of order $\vert c_i\vert_{\max}\lesssim O(10)$, satisfy the proton decay constraint (\ref{MXprotonconstraint}) easily, and are also fairly safe against the expected ten-year improvements in the bound. For example, of the models with one Higgs multiplet (see Table \ref{results1opSU5table}), only the ${\bf 200}$ model (and possibly the ${\bf75}$ model, depending on the choice of $M_{Pl}$) feature natural gauge coupling unification, with the unification scale $5\times10^{18}\,\text{GeV}$ (resp.~$8\times10^{15}\,\text{GeV}$) above the bound (\ref{MXprotonconstraint}) in either case. This also holds for models with two Higgs multiplets, see Figs.~\ref{SU5w24and75-cmax-epsilons} and \ref{SU5w24and200_w75and200}: Under the condition $\vert c_i\vert_{\max}<5$, the lowest unification scale $M_X=1.4\times10^{16}\,\text{GeV}$ is here achieved in the model with a ${\bf75}$ and a ${\bf200}$ Higgs (Fig.~\ref{SU5w24and200_w75and200}, right panel) when assuming $\xi=\xi^\text{run}_\text{red}=13.6$; the other two models with two multiplets automatically have $M_X>3\times10^{16}\,\text{GeV}$ if gauge coupling unification with Wilson coefficients of this size is required.

Quite generally, any reasonable non-supersymmetric unification model featuring natural gauge coupling unification via (\ref{unificationcondition}) satisfies the proton decay constraint (\ref{MXprotonconstraint}). This can be seen from the general estimate (\ref{estimatewithnumerics}) of the parameters necessary for gauge coupling unification:
\begin{equation}\label{modelsatisfyprotongenerally}
M_X~\gtrsim~\frac{O(0.1)}{\xi\sqrt{h}}\,\frac{1.2\times10^{19}\,\text{GeV}}{\max_i\vert c_i\vert}~\gtrsim~\frac{0.1}{15\sqrt{4}}\,\frac{1.2\times10^{19}\,\text{GeV}}{10}~=~4\times10^{15}\,\text{GeV}~,
\end{equation}
so that (\ref{MXprotonconstraint}) still holds despite all the very conservative parameter choices in (\ref{modelsatisfyprotongenerally}) which are by no means necessary or particularly desirable. Models with more natural parameters (i.e., with smaller $\vert c_i\vert_{\max}$, reasonable $h$, and possibly choosing $\xi_\text{red}=\sqrt{8\pi}$) satisfy the proton lifetime constraint very easily.

The intuitive reason why our models naturally obey the proton decay constraint so readily is clear: In non-supersymmetric models, the couplings $\alpha_s(\mu)$ miss each other by quite a bit, so the modifications to the unification condition (\ref{unificationcondition}) need to be relatively sizeable $\epsilon_s\sim O(0.1)$ in order to achieve exact gauge coupling unification; then, since we require naturalness $\vert c_i\vert_{\max}\lesssim O(10)$, these sizes $\epsilon_s\sim c_iM_X/M_{Pl}$ of the effective gravitational corrections (\ref{noncanonicalterms}) must be due mainly to the proximity of the Planck scale $M_{Pl}=1.2\times10^{19}\,\text{GeV}/\xi$ to the unification scale $M_X$ and, consequently, to the gauge boson masses (\ref{gbmassrequirement}). Typically, such heavy gauge bosons ensure the proton decay constraint (\ref{MXprotonconstraint}).

On the other hand, whereas the natural models easily satisfy current proton decay limits, there are situations conceivable in which the constraints in the not too distant future will come close to testing some of the models, either excluding them or, if the proton is actually seen to decay, strongly restricting their parameter space. For example, a model with Higgs content ${\bf24}$, ${\bf75}$ and ${\bf200}$ can achieve unification at $M_X=8\times10^{15}\,\text{GeV}$, the projected constraint in less than ten years from now \cite{protondecay10yearprojection}, if merely the reduced Planck scale ($\xi=\xi_\text{red}=\sqrt{8\pi}\approx5$) and somewhat large coefficients $\vert c_i\vert_{\max}\sim20$ are accepted (see Fig.~\ref{general3estimatefigure}). Alternatively, the smallest model with two Higgs multiplets ${\bf24}$ and ${\bf75}$ can reach this limit when taking $\xi=\xi_\text{red}^\text{run}=8$ and allowing $\vert c_i\vert_{\max}\sim15$, and similarly can the other models with two multiplets (in each example, only the product $\xi\vert c_i\vert_{\max}$ is fixed whereas the sizes of both factors can be traded back and forth). Regarding the exact numerical values in this discussion, one has to keep in mind that there is some uncertainty in the exact numerical $O(1)$ factor in (\ref{protondecaytime}), although it enters the final bound (\ref{MXprotonconstraint}) only to the power of $1/4$, and in higher-order and threshold corrections to the gauge coupling running.

We have seen that the natural models with gauge coupling unification through (\ref{unificationcondition}) are physically viable as they automatically evade the proton lifetime constraint that naively excludes non-supersymmetric unification. However, some of the less natural unification models will come close to the proton limit in the near future, which will exclude some of them, or constrain their parameter space drastically if proton decay is observed.

\section{Simultaneous unification of gauge and gravitational interactions}\label{gaugegravitysection}
As can be seen from Figs.~\ref{SU5w24and75-cmax-epsilons}--\ref{general3estimatefigure} or from estimate (\ref{estimatewithnumerics}), gauge coupling unification at large $M_X\sim M_{Pl}$ is possible through mechanism (\ref{unificationcondition}) naturally, i.e.~with small Wilson coefficients $c_i$. Unification of the three standard model gauge interactions with each other on the one hand and with the gravitational interaction on the other hand at the same scale (``gauge-gravity unification''; cf.~\cite{Agashe:2002pr,MPlUnification}) therefore constitutes another scenario naturally achievable through the modified unification condition (\ref{unificationcondition}). However, one should keep in mind that our analysis is based on an effective theory approach and we should therefore consider carefully the expansion we are using. If the operators of dimension 5 and higher that we are considering are of non-perturbative nature, then the expansion is in powers of $1/M_{Pl}$ and one may worry that it could break down once the Higgs vevs get close to the Planck mass $M_{Pl}$. In any case, it is an interesting numerical coincidence that the corrections to the unification condition allow to shift the unification scale close to the Planck scale for a natural set of parameters.

Approximate parameter values $c_i$ of such gauge-gravity unification models can be read off from estimate (\ref{estimatewithnumerics}) by setting $M_X=M_{Pl}/O(1)$:
\begin{equation}\label{estimateforgaugegravity}
\max_i\vert c_i\vert~\approx~O(1)\,\frac{O(0.2\to1)}{\sqrt{h}}~,
\end{equation}
where the $O(0.2\to1)$ estimate stems from the $M_X\approx M_{Pl}$ range of numerical studies like Fig.~\ref{general3estimatefigure} (right panel) and $h$ denotes the number of Higgs multiplets at the grand unification scale. This suggests the possibility of very natural gauge coupling unification at or near $M_{Pl}$ and is already apparent in the only model with one Higgs multiplet that achieves unification close to the Planck scale: The ${\bf200}$ model has $M_X=5.2\times10^{18}\,\text{GeV}$, which is related to the Planck scale by $M_X=0.4\xi M_{Pl}$, and the required Wilson coefficient is natural $c=0.53/\xi$ (see Table \ref{results1opSU5table}). As another example, gauge coupling unification can be achieved for appropriate parameter choices in the model with a ${\bf24}$ and a ${\bf75}$ (Fig.~\ref{SU5w24and75-cmax-epsilons}) at the scale $M_X=M_{Pl}=1.2\times10^{19}\,\text{GeV}/\xi$ for any of the reasonable exemplary choices $\xi=1$, $\xi_\text{red}=\sqrt{8\pi}$, $\xi^\text{run}=1.6$ or $\xi^\text{run}_\text{red}=8.0$; the corresponding model parameters $\vert c_i\vert_\text{max}=0.23$, $0.21$, $0.23$ and $0.20$ are all natural and of almost equal sizes independent of $\xi$, see (\ref{estimateforgaugegravity}). Similar numerical estimates hold for the other two- and three-Higgs models. In these models, any $O(1)$ factor in the relation $M_X=M_{Pl}/O(1)$ can be easily accommodated as well by corresponding $O(1)$ changes to the Wilson coefficients, whereas this is not possible for the sole-${\bf200}$ model as it only permits one discrete unification scale (Table \ref{results1opSU5table}).

In the effective field theory spirit of Section \ref{sectionSU5setup}, operators of dimension higher than 5 are also present, e.g.~higher-dimensional generalizations of (\ref{dim5operators}):
\begin{equation}\label{dim67operators}
{\cal L}=\frac{c_6}{4M_{Pl}^2}H_1H_2G_{\mu\nu}G^{\mu\nu}+\frac{c_7}{4M_{Pl}^3}H_1H_2H_3G_{\mu\nu}G^{\mu\nu}+\ldots~.
\end{equation}
After the Higgs multiplets acquire vevs at the scale $M_X=M_{Pl}/O(1)$, they contribute to the gauge kinetic terms (\ref{noncanonicalterms}) as well:
\begin{equation}
{\cal L}=\sum_{s=1}^3-\frac{1}{4}\left(1+\epsilon_s+\frac{c_6\delta^{(6)}_s}{g_G^2\,O(1)^2}+\frac{c_7\delta^{(7)}_s}{g_G^3\,O(1)^3}+\ldots\right)F_{(s)\mu\nu}^aF_{(s)}^{a\mu\nu}~,
\end{equation}
with the corrections $\epsilon_s\sim c_5\delta_sM_X/g_GM_{Pl}$ from the dimension-5 operators (\ref{defepsilon}). Depending on the group theory factors $\delta^{(6,7)}_s$ (analogous to the $\delta_s$ in Table \ref{deltaSU5table}) and on the constant in the relation $M_X=M_{Pl}/O(1)$, this expansion might or might not be controlled perturbatively. If it is 
not, one cannot claim perturbative gauge-gravity unification at the Planck scale. Nevertheless, the fact that mechanism (\ref{unificationcondition}) allows in principle to naturally adjust the unification scale to a high scale $\sim M_{Pl}$ might at least be taken as a hint that gauge-gravity unification is a possible scenario, even if the necessary parameter values or the last piece of the evolution cannot be computed perturbatively.

\section{The non-supersymmetric ${\bf{SO(10)}}$ case}\label{sectionSO10}
In this section we give an account of the effects from the dimension-5 operators (\ref{dim5operators}) in models with grand unified group $G=SO(10)$. We will describe the setup for $SO(10)$ by emphasizing the differences to the $SU(5)$ case described in Section \ref{sectionSU5setup} and give a few numerical unification results similar to Section \ref{sectionSU5results}. We find that the effects from Sections \ref{protondecaysection} (on proton decay) and \ref{gaugegravitysection} (on gauge-gravity unification) can occur with similar sizes for $SO(10)$ as well, although the $SO(10)$ formalism is more general than the $SU(5)$ one; in particular, a continuously variable unification scale $M_X$ can now be achieved with only one single dimension-5 operator. For this reason, and for the beauty of $SO(10)$ unification, we find this treatment worthwhile.

For $G=SO(10)$, there are \emph{two} inequivalent ways to embed the standard model group $G_{321}=SU(3)_C\times SU(2)_L\times U(1)_Y$ into $G$ consistent with the charge assignments of the standard model fermions: the ``normal embedding'' $G_{321}\subset SU(5)\subset SO(10)$ \cite{Fritzsch:1974nn}, and the ``flipped embedding'' $G_{321}\subset SU(5)\times U(1)_X\subset SO(10)$ with $G_{321}\nsubseteq SU(5)$ \cite{Barr:1981qv} (see the Appendix for a careful treatment of the $SO(10)$ group theory). Statements without qualifier in the following apply to either embedding. Note, nowhere are we implying that either $SU(5)$ or $SU(5)\times U(1)_X$ (or any other subgroup) be intermediate unbroken symmetries at any scale.

In the $SO(10)$ grand unified theory, the gravitationally induced effective dimension-5 operators (\ref{dim5operators}) can be formed with Higgs multiplets $H_i$ in any of the four irreps $r_i={\bf 1}$, ${\bf 54}$, ${\bf 210}$, ${\bf 770}$. However, contrary to the $SU(5)$ case, the vev $\langle H_i\rangle$ is not necessarily uniquely specified (up to normalization) by requiring it to be invariant under the standard model group $G_{321}$; rather, this $G_{321}$-invariance merely restricts to a 3-\ resp.~4-dimensional subspace inside the ${\bf210}$ resp.~${\bf770}$ irreps, so that the length as well as the direction of the vev has to be specified for Higgses in these irreps. This will have the important consequence that with merely \emph{one} dimension-5 operator (\ref{dim5operators}), built with either a ${\bf210}$ or ${\bf770}$ Higgs, one can continuously vary the ratio among the $\epsilon_s$ in the unification condition (\ref{unificationcondition}) by continuous variation of the vev direction, ultimately leading to a continuously variable unification scale $M_X$; the $SU(5)$ case could achieve this continuity only with at least two Higgs multiplets.

To parametrize these vev directions for calculations, one has to specify a basis in these 3-\ or 4-dimensional subspaces. Two possible, distinct choices of this basis are given in the Appendix: the vectors in the first resp.~second basis have definite transformation properties under the $SU(5)\times U(1)_X$ resp.~$SU(4)_C\times SU(2)_L\times SU(2)_R$ maximal subgroups of $SO(10)$ (see the first two columns of Tables \ref{SO10entriesSU5U1X} resp.~\ref{SO10entriesSU4CSU2R} for these transformation properties). We choose to classify according to these two subgroups since they are the only maximal subgroups that can occur as intermediate symmetries in $SO(10)\to G_{321}$ breaking consistent with the standard model charge assignments. These two choices therefore readily facilitate the analysis of $SO(10)$ breaking with intermediate gauge symmetries, see below (the absence of intermediate symmetries is assumed for now).

With such a chosen basis, the modifications $\epsilon_s$ to the three standard model gauge kinetic terms (\ref{noncanonicalterms}) are now, instead of (\ref{defepsilon}), double sums (\ref{defepsilonAPP})
\begin{equation}\label{defepsilondouble}
\epsilon_s=\sum_i\frac{c_i}{M_{Pl}}\sum_jv_{(i)j}\delta^{(i)j}_{s}~~~~(\text{for}~s=3,2,1)~,
\end{equation}
where the second sum runs over the directions $j=1,\ldots,3\,\,(4)$ if $H_i$ is in the ${\bf210}$ (${\bf770}$) representation (for the ${\bf1}$ and ${\bf54}$ irreps, $j=1$ only). The Clebsch-Gordan coefficients $\delta^{(i)j}_s$ depend on the embedding $G_{321}\subset SO(10)$ (normal or flipped), on the representation $r_i$ of the Higgs $H_i$ and on the choice of basis vectors $j$ in the standard model singlet subspace of the respective representation. These $\delta^{(i)j}_s$ can all be taken from Table \ref{SO10entriesSU5U1X} (for basis vectors with definite $SU(5)\times U(1)_X$ transformation properties) and Table \ref{SO10entriesSU4CSU2R} (definite $SU(4)_C\times SU(2)_L\times SU(2)_R$ transformations) via (\ref{defepsilonAPP}).

One other difference between the $SO(10)$ and the $SU(5)$ cases is the fact that none of the ${\bf54}$, ${\bf210}$ or ${\bf770}$ multiplets can give mass to all of the non-standard model (``superheavy'') gauge bosons and that, furthermore, the masses of the gauge bosons which actually do acquire mass from these multiplets are generally unequal. Therefore, both the statement from the $SU(5)$ case that the mass $M_\text{gb}$ of the superheavy gauge bosons may originate solely from Higgses involved in dimension-5 operators (\ref{dim5operators}) and the requirement (\ref{gbmassrequirement}) that this mass be close (or equal) to the unification scale $M_X$ have to be amended. First, additional Higgs multiplets in other irreps (typically ${\bf45}$, ${\bf16}$, or ${\bf126}$; see also Table \ref{etatable}) are necessary to break the $SO(10)$ gauge symmetry down to the standard model $G_{321}$ and they contribute to the masses of the superheavy gauge bosons as well. And for definiteness, we now assume that the Higgs multiplets involved in dimension-5 operators (\ref{dim5operators}) account for half of the average squared superheavy gauge boson masses, and we require that this \emph{averaged} superheavy mass now be equal to the unification scale, see (\ref{assertMX}) in the Appendix. The necessity for some such choice causes uncertainty in actual numerical unification calculations greater than in the $SU(5)$ case.

As mentioned above, continuously variable unification scales $M_X$ can be achieved in $SO(10)$ models with a single Higgs in either the ${\bf210}$ or the ${\bf770}$ representation, since the direction $\{v_{(i=1)j}\}_{j=1,\ldots,3\,(4)}$ of their vev can be varied continuously in (\ref{defepsilondouble}). Minimizing the Higgs content and the number of dimension-5 operators in this way seems attractive. Further note, that a ${\bf770}$ causes the strong coupling regime of the grand unified theory to be quite close to the unification scale (Table \ref{etatable}), although, depending on the gap between the unification and the Planck scale, the theory may still be perturbative up to the onset of quantum gravity.

\begin{figure}
\includegraphics[scale=0.8]{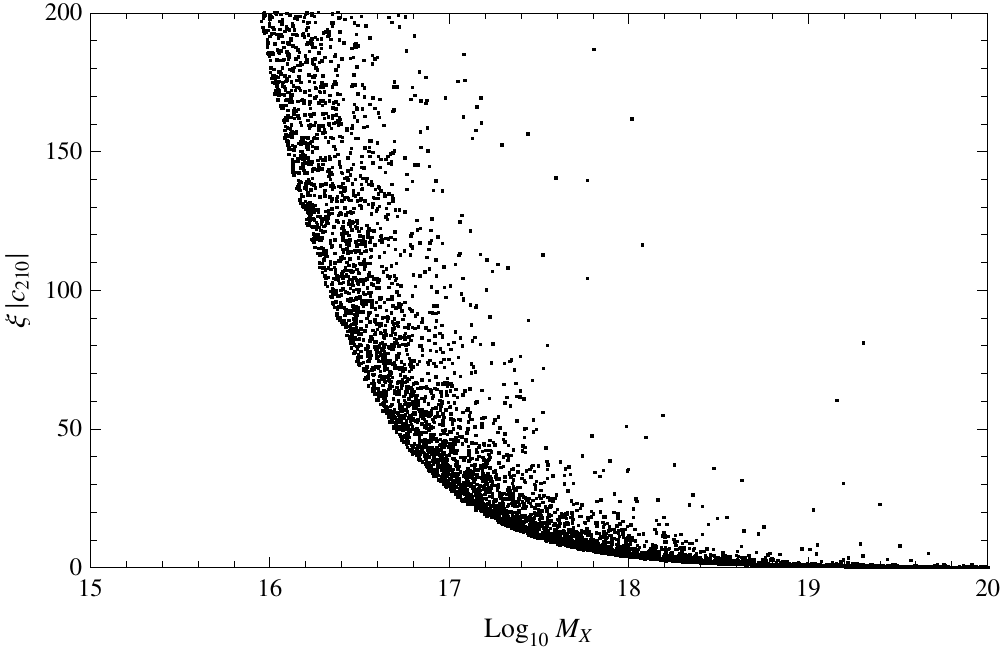}~~~~~~~~~~\includegraphics[scale=0.8]{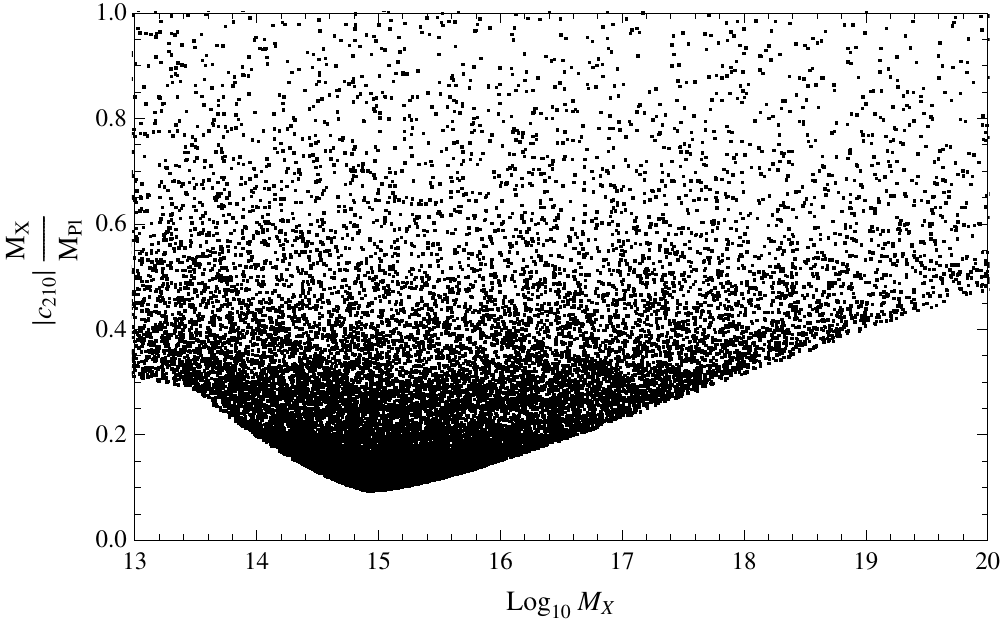}
\caption{For the $SO(10)$ model with a ${\bf210}$ Higgs, a random sample is taken of $30,000$ points $\{c_{\bf210},v_{({\bf210})j}\}$ of the parameter space which yield gauge coupling unification, cf.~Fig.~\ref{general3estimatefigure}. Estimate (\ref{estimatewithnumerics}) is valid here as well (see right panel), and numerical gauge coupling unification at any $10^{13}\,\text{GeV}<M_X<10^{20}\,\text{GeV}$ can be achieved with the $O(0.1)$ constant in (\ref{estimatewithnumerics}) being at most $0.5$ (the caveat from Fig.~\ref{general3estimatefigure} applies here as well).\label{example210inSO10figure}}
\end{figure}

As a numerical example, Fig.~\ref{example210inSO10figure} shows gauge coupling unification results in the $SO(10)$ model with a single ${\bf210}$ Higgs (identical results obtain for both the normal and the flipped embedding due to the orthogonal relation (\ref{orthogrel})), similar to Fig.~\ref{general3estimatefigure} for $SU(5)$ with three Higgses. The lower bound (\ref{estimatewithnumerics}) holds here as well, even with the same factor $O(0.1)$, and unification \emph{can} be achieved for any $M_X$ in the displayed range if this factor is allowed to be $0.5$ (see right panel). Therefore, unification is naturally safe from the proton lifetime limit in $SO(10)$ as well, but can also come close to it for some points in the parameter space, and unification near or at the Planck scale is achievable in a very natural way, both similar to the corresponding $SU(5)$ scenarios described in detail in Sections \ref{protondecaysection} and \ref{gaugegravitysection}.

\bigskip

In a scenario with several breaking scales, if the intermediate gauge symmetry is $SU(5)\times U(1)_X$ (or merely $SU(5)$) or the Pati-Salam group $SU(4)_C\times SU(2)_L\times SU(2)_R$, respectively, the directions of both the ${\bf210}$ and ${\bf770}$ vevs at the unification scale $M_X$ are fixed, as these vevs have to be singlets under the intermediate group, which allows only the $\delta^{(i)j}_s$ from rows 1, 3, 6 in Table \ref{SO10entriesSU5U1X} resp.~rows 1, 2, 3, 6 in Table \ref{SO10entriesSU4CSU2R}. Therefore, similar to the case of an $SU(5)$ unified group, only linear combinations of the effects from at least two operators (\ref{dim5operators}) can achieve continuously varying $M_X$. The effect of a dimension-5 operator with a Higgs that acquires its vev at an intermediate scale $M_I<M_X$ is suppressed relatively by $\sim M_I/M_X$. Note, since there are no singlet vevs for a ${\bf54}$ Higgs under the $SU(5)$ intermediate group, it can acquire nonzero vev only at the final breaking scale to the standard model group $G_{321}$ and is therefore almost irrelevant for our effect in the case of an $SU(5)$ intermediate gauge symmetry.

In several previous analyses of the $SO(10)$ case (e.g., \cite{Chakrabortty:2008zk}), only Higgs vevs that are invariant under intermediate gauge groups larger than $G_{321}$ have been considered (namely under the $SU(5)$ Georgi-Glashow or the $SU(4)_C\times SU(2)_L\times SU(2)_R$ Pati-Salam subgroups), thereby omitting most of the possibilities discussed above for $SO(10)$ breaking without intermediate symmetries, i.e.~the continuous variability of $M_X$ with merely one dimension-5 operator.

Exactly analogous effects obtain for other grand unified gauge groups, like $E_6$, and they are even of similar size as in $SU(5)$ and $SO(10)$ \cite{Chakrabortty:2008zk,Martin:2009ad}. Furthermore, the formalism necessary for the $SO(10)$ treatment is already the most general for any unification group, so we leave the analysis of gauge coupling unification through the mechanism of dimension-5 operators (\ref{dim5operators}) at the $SU(5)$ and $SO(10)$ cases discussed so far.

\section{Comparison to the supersymmetric case}\label{susysection}
The influence of the dimension-5 operators (\ref{dim5operators}) has been studied \cite{Hill:1983xh} in models of supersymmetric grand unification \cite{susyguts,Jones:1982,Marciano:1981} as well, although most studies have focused on the effect of the supersymmetrized version of (\ref{dim5operators}) in the creation of non-universal gaugino masses in a scenario where the UV completion is a ${\cal N}=1$ supergravity theory, see \cite{Martin:2009ad} and references therein. The group theory needed to calculate the resulting gaugino mass ratios is very similar to the formalism for obtaining the modified unification condition (\ref{unificationcondition}) \cite{Ellis:1985jn,Drees:1985bx}; see the Appendix for the detailed $SU(5)$ and $SO(10)$ group theory.

The effect of these dimension-5 operators on the unification of gauge couplings in supersymmetric theories and on the unification scale has been noted before \cite{Hill:1983xh,Ellis:1985jn,Drees:1985bx,Huitu:1999eh,ourprl,Chakrabortty:2008zk}. In some of these works, however, the low-energy inputs $\alpha_s(m_Z)$ or the supersymmetry breaking scale $m_{SUSY}$ (above which the $\beta$--function coefficients (\ref{betafunctioncoeffsAPP}) are taken to be $(b_1,b_2,b_3)=(33/5,1,-3)$) have been treated as uncertain parameters, and mostly only the effect from a single dimension-5 operator had been taken into account (except in \cite{Huitu:1999eh}).

Fixing the low-energy gauge coupling values (\ref{initialvaluesAPP}) and $m_{SUSY}=1\,\text{TeV}$ \cite{Amaldi:1991cn}, as roughly required to solve the hierarchy problem of the standard model, one obtains for supersymmetric $SU(5)$ models with one dimension-5 operator the unification possibilities in Table \ref{results1opSUSYSU5table} (at one loop), in analogy with Table \ref{results1opSU5table} for non-supersymmetric $SU(5)$ (note, supersymmetric $SU(5)$ with a ${\bf200}$ Higgs enters strong coupling shortly above the unification scale, cf.~Table \ref{etatable}).

\begin{table}[htb]
\begin{center}
\begin{tabular}{c||c|c|c|c|c||c|c|c|c}
\hline
$H$ irrep & $M_X/\text{GeV}$ & $1/\alpha_G$ & $c$ & $v/\text{GeV}$ & $\text{max}_s\,\vert\epsilon_s\vert$ & $\xi=1$ & $\xi_\text{red}=\sqrt{8\pi}$ & $\xi^\text{run}$ & $\xi^\text{run}_\text{red}$ \\
\hline
\hline
${\bf 1}$ & \multicolumn{5}{c||}{unification by mechanism (\ref{unificationcondition}) impossible} \\
\hline
${\bf 24}$ & $1.1\times10^{16}$ & $25.9$ & $31.1/\xi$ & $2.4\times10^{16}$ & $0.024$ & $1$ & $5$ & $2.3$ & $11.6$ \\
\hline
${\bf 75}$ & $3.2\times10^{16}$ & $25.8$ & $-12.0/\xi$ & $5.6\times10^{16}$ & $0.033$ & $1$ & $5$ & $3.1$ & $15.4$ \\
\hline
${\bf 200}$ & $8.5\times10^{16}$ & $26.8$ & $13.1/\xi$ & $1.2\times10^{17}$ & $0.105$ & $1$ & $5$ & $4.4$ & $22.1$ \\
\hline
\end{tabular}
\end{center}
\caption{This table shows the parameters that a SUSY-$SU(5)$ model must have if, at one loop, unification happens by means of mechanism (\ref{unificationcondition}) with only one dimension-5 operator (\ref{dim5operators}). The sizes of the required Wilson coefficients $c$ depend on the chosen suppression scale (\ref{suppressionscale}), parametrized by $\xi$.\label{results1opSUSYSU5table}}
\end{table}

\begin{figure}
\includegraphics[scale=0.8]{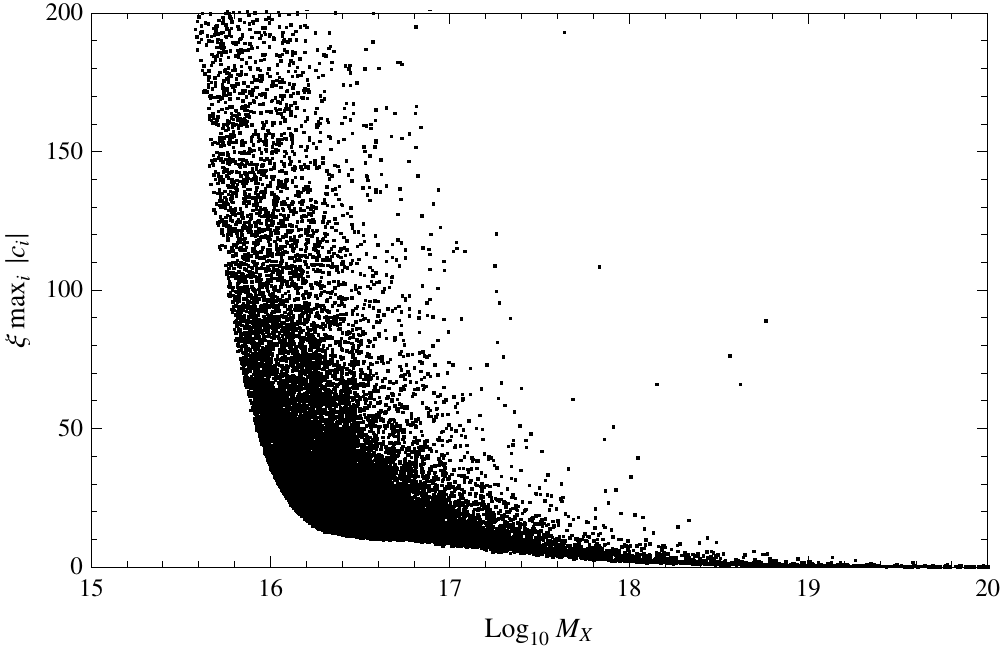}~~~~~~~~~~\includegraphics[scale=0.8]{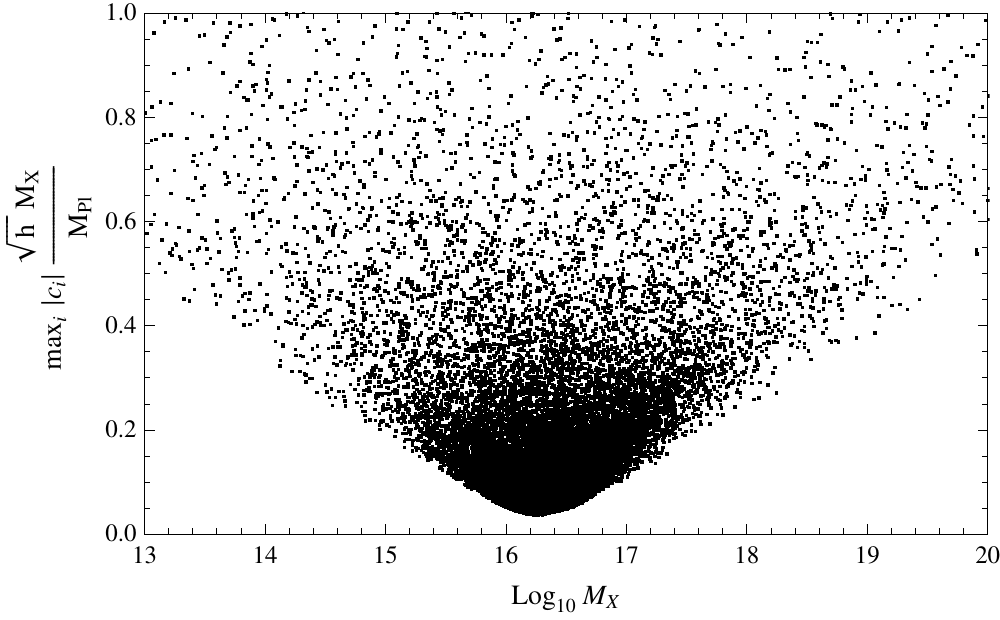}
\caption{Similar to Fig.~\ref{general3estimatefigure}, but for the supersymmetric $SU(5)$ model with the three ($h=3$) Higgs multiplets ${\bf24}$, ${\bf75}$ and ${\bf200}$. Again, numerical gauge coupling unification at any $10^{13}\,\text{GeV}<M_X<10^{20}\,\text{GeV}$ can be achieved by choosing $0.5$ as the constant in (\ref{estimatewithnumerics}) (but cf.~caveat from Fig.~\ref{general3estimatefigure}). Most of the randomly chosen unification points lead to unification scales $M_X\sim2\times10^{16}\,\text{GeV}$, where, in supersymmetric models, the standard model gauge couplings almost meet.\label{SUSYgeneral3estimatefigure}}
\end{figure}

For a general number of Higgs fields, the situation is shown in Fig.~\ref{SUSYgeneral3estimatefigure}, in analogy to Fig.~\ref{general3estimatefigure}. As can be seen, exact supersymmetric gauge coupling unification around $2\times10^{16}\,\text{GeV}$ can be achieved with natural-sized dimension-5 operators (\ref{dim5operators}), which is expected as this is where the standard model gauge couplings come closest to each other in supersymmetric models; but higher unification scales are also possible in a natural way with certain values of the coefficients $c_i$. Note that the supersymmetric unification model from Fig.~\ref{SUSYgeneral3estimatefigure} has a Landau pole roughly half an order of magnitude above the unification scale $M_X$, cf.~Table \ref{etatable}.

\section{Summary and Conclusion}\label{conclusionsection}
We have studied the effects from multiple gravitationally induced dimension-5 interactions $c_iH_iG_{\mu\nu}G^{\mu\nu}/4M_{Pl}$ (\ref{dim5operators}), naturally present in any grand unified theory, on the unification of gauge couplings. These operators can modify the canonical gauge kinetic terms (\ref{noncanonicalterms}) and the coupling unification condition (\ref{unificationcondition}) by up to $\epsilon_s\sim c_i\langle H_i\rangle/M_{Pl}\sim c_iM_X/g_GM_{Pl}\sim0.1\%-30\%$ (\ref{defepsilon}) after the Higgs multiplets acquire vevs in grand unified symmetry breaking. The size of the effect depends on the unification scale $M_X$, the Wilson coefficients $0.1\lesssim\vert c_i\vert\lesssim10$, and the suppression scale $5\times10^{17}\,\text{GeV}\lesssim M_{Pl}\lesssim1.2\times10^{19}\,\text{GeV}$ (\ref{suppressionscale}).

Modifications $\epsilon_s$ of this size can effect perfect gauge coupling unification, at scales $M_X$ significantly different than naively expected, as the standard model gauge couplings $\alpha_s(\mu)$ ($s=1,2,3$) differ from each other by at most $\lesssim50\%$ in the wide range $10^{13}\,\text{GeV}<\mu<10^{19}\,\text{GeV}$, both in the case with and without supersymmetry. When two or more dimension-5 operators are present in a theory (or a single ${\bf210}$ or ${\bf770}$ in $SO(10)$, see Section \ref{sectionSO10}), then the scale $M_X$ of gauge coupling unification can be varied in a continuous and controlled way as a function of the Wilson coefficients $c_i$ and Higgs vevs $\langle H_i\rangle$, which are practically model parameters. We have focused on non-supersymmetric $SU(5)$ and $SO(10)$ theories, for which grand unification seemed, previously, to be difficult \cite{Amaldi:1991cn}, and we find that gauge coupling unification is possible, in a natural way, at any $M_X\sim10^{17}\,\text{GeV}-1.2\times10^{19}\,\text{GeV}$. For illustration we now give a few numerical unification results (all non-supersymmetric models):

\renewcommand{\theenumi}{(\alph{enumi})}\renewcommand{\labelenumi}{\theenumi}
\begin{enumerate}
\item\label{exampleSU52475}$SU(5)$ with ${\bf24}$ and ${\bf75}$ Higgses, see Fig.~\ref{SU5w24and75-cmax-epsilons}: Perfect gauge coupling unification at $M_X=10^{17}\,\text{GeV}$ happens for some choice of parameter values $c_{24},\,c_{75}$ with $\vert c_{i}\vert\lesssim3$, if $M_{Pl}=1.2\times10^{19}\,\text{GeV}/\sqrt{8\pi}=2.4\times10^{18}\,\text{GeV}$ is assumed; the modifications $\epsilon_s$ to the gauge kinetic terms are all less than $15\%$.
\item\label{exampleSU52475200}$SU(5)$ with ${\bf24}$, ${\bf75}$ and ${\bf200}$, Fig.~\ref{general3estimatefigure} and estimate (\ref{estimatewithnumerics}): Unification at any $M_X>10^{18}\,\text{GeV}$ is possible for several choices of $c_{24}$, $c_{75}$, $c_{200}$ with $\vert c_i\vert<2$, when $M_{Pl}=1.2\times10^{19}\,\text{GeV}$ is assumed. In fact, the bigger a unification scale one wants to achieve, the smaller Wilson coefficients are required: $\max_i\vert c_i\vert\approx2\cdot(10^{18}\,\text{GeV}/M_X)$; analogous statements hold (at fixed $M_{Pl}$) in all examples \ref{exampleSU52475}--\ref{exampleSO10210} with continuous variability of $M_X$, cf.~(\ref{estimatewithnumerics}).
\item\label{exampleSO10210}$SO(10)$ with a single ${\bf210}$, Fig.~\ref{example210inSO10figure}: Exact unification at $M_X=3\times10^{17}\,\text{GeV}$ is possible with $c_{210}\approx2.5$ for a certain (continuous) set of directions of the Higgs vev $\langle H^{ab}\rangle$, if $M_{Pl}=1.2\times10^{19}\,\text{GeV}/\sqrt{8\pi}=2.4\times10^{18}\,\text{GeV}$ is assumed.
\item\label{exampleSU52475gaugegravity}$SU(5)$ with ${\bf24}$ and ${\bf75}$, Fig.~\ref{SU5w24and75-cmax-epsilons} and equation (\ref{estimateforgaugegravity}): Numerical gauge coupling unification at the very high scale $M_X=M_{Pl}$, i.e., at the Planck scale, is possible for some choice of Wilson coefficients with $\vert c_{24}\vert,\,\vert c_{75}\vert<0.25$ (irrespective of the specific choice of $M_{Pl}$).
\item\label{exampleSU5200}$SU(5)$ with a single ${\bf200}$, Table \ref{results1opSU5table}: Exact gauge-coupling unification can happen, but only for a discrete choice $c_{200}=0.5$ \cite{Chakrabortty:2008zk} (here, $M_{Pl}=1.2\times10^{19}\,\text{GeV}$ is assumed); in this case, the unification scale $M_X=5.2\times10^{18}\,\text{GeV}$ cannot be shifted continuously.
\end{enumerate}

We have shown (Section \ref{protondecaysection}) that non-supersymmetric unification, achieved via these dimension-5 operators, is not in conflict with the current bound on the proton lifetime (\ref{MXprotonconstraint}), since, with natural-sized Wilson coefficients $c_i$, only unification scales $M_X\gtrsim10^{16}\,\text{GeV}$ are reasonably possible, cf.~(\ref{modelsatisfyprotongenerally}). Consequently, experimental improvements of the proton decay bound will constrain our models only weakly within the foreseeable future. Our main conclusion is that fairly minimal models of non-supersymmetric $SU(5)$ and $SO(10)$ unification are easily viable through the mechanism described. In particular, supersymmetry does not have to be invoked to save the idea of grand unification.

We also note that, with two or more dimension-5 operators in the theory, gauge coupling unification at or near the Planck scale $M_X\sim M_{Pl}$ can happen for very natural parameter choices $c_i\sim O(0.1\to1)$, cf.~(\ref{estimateforgaugegravity}) in Section \ref{gaugegravitysection}. Of course, our approach cannot be fully justified very near the Planck scale, due to incalculable corrections from quantum gravity. But the fact that one can, by continuous variation of the Wilson coefficients $c_i$ in a natural domain, push the unification scale $M_X$ close towards the Planck scale, may be suggestive of a scenario of simultaneous gauge-gravity unification.

As we have seen, when unification happens according to the mechanism presented in this paper, then unification scales $M_X$ much larger than usually assumed are favored, in the sense that probably only those are achievable in a natural way. Thus, the mechanism here is distinct from some other known unification scenarios \cite{Lavoura:1993su,Weinberg:1980wa}. The difference lies in the fact that the importance of the dimension-5 operators (\ref{dim5operators}) increases for unification scales $M_X$ closer to the Planck scale (implying larger Higgs vevs $\langle H\rangle$), which causes a bigger modification to the unification condition (\ref{unificationcondition}) and makes this larger $M_X$ consistent. But, for example, two-loop and threshold effects are not significantly enhanced at larger unification scales, since they originate from within the grand unified theory, rather than from intrinsically shorter distance effects like strong quantum gravity.

\bigskip
\bigskip

\emph{Acknowledgments}~---~SDHH and DR are supported by the Department of Energy under DE-FG02-96ER40969. The authors thank Stephen Barr and Tuhin Roy for discussion.

\bigskip

\appendix*\section{Normalization conventions and group theory}
In this appendix, we give a self-contained account of the group theoretical aspects and normalization details that were omitted in the main text.

Our first goal is to establish conventions for the quantities appearing in the dimension-5 operators (\ref{dim5operators}) of interest, aiming towards criteria for assessing the individual size and importance of each operator $i$ based on its Wilson coefficient $c_i$, and also the relative sizes of different operators (\ref{dim5operators}) appearing in the same Lagrangian. We intend to establish normalization conventions in such a way that operators with Wilson coefficients $\vert c_i\vert\sim1$ are expected in the effective Lagrangian of a grand unified theory after integrating out gravitational interactions, which then helps judge the naturalness of a particular model, i.e.~its resonableness: If $\vert c_i\vert\gg1$ (for any $i$), the effective Lagrangian might not be a good low-energy description and one might lose perturbative control, whereas $\vert c_i\vert\ll1$ constitutes fine-tuning, as setting $c=0$ does not enhance the symmetry \cite{tHooft:1979bh}. Note, the calculations and numerical gauge coupling unification results in this paper are valid, at face value, regardless of such criteria.

The effective operators $\sim HG_{\mu\nu}G^{\mu\nu}$ we are considering are formed from two gauge field strengths $G_{\mu\nu}$, transforming in the adjoint representation $\bf{G}$ of the grand unified gauge group $G$, and one Higgs multiplet $H$, transforming in an irreducible representation (irrep). The contraction $HG_{\mu\nu}G^{\mu\nu}$ can yield a gauge singlet only if $H$ transforms in an irrep contained in the (conjugate of the) symmetric product $\left({\bf G}\otimes{\bf G}\right)_s$ of two adjoint representations of the gauge group. For $SU(5)$ and $SO(10)$ unified gauge groups $G$, these possible irreps are the direct summands in the following decompositions (for this and other group theory facts, see \cite{Slansky:1981yr}):
\begin{eqnarray}\label{higgsirreps}
G=SU(5),~{\bf G}={\bf 24}:&&\left({\bf 24}\otimes{\bf 24}\right)_s={\bf 1}\oplus{\bf 24}\oplus{\bf 75}\oplus{\bf 200}~;\\
G=SO(10),~{\bf G}={\bf 45}:&&\left({\bf 45}\otimes{\bf 45}\right)_s={\bf 1}\oplus{\bf 54}\oplus{\bf 210}\oplus{\bf 770}~.\nonumber
\end{eqnarray}
(In this paper, the ${\bf 210}$ of $SO(10)$ will always denote the 210-dimensional irrep in the same congruency class as the ${\bf 54}$ and ${\bf 770}$, namely in the congruency class $0$, to which also the adjoint ${\bf G}={\bf 45}$ belongs.) The irreps on the right-hand sides of (\ref{higgsirreps}) are all real representations, as the adjoint ${\bf G}$ is real itself.

Since they are contained in $\left({\bf G}\otimes{\bf G}\right)_s$, each multiplet $H$ in an irrep (\ref{higgsirreps}) can be written in component form as $H^{ab}$ with symmetric indices $a$, $b$ of the respective adjoint representation ${\bf G}$. Under an infinitesimal gauge transformation $(1+i\alpha^cT^c)$ these components then transform according to
\begin{equation}\label{GtimesGtrafo}
H^{ab}~\to~H^{ab}+i\alpha^c\left((t^c_{\bf G})^{aa'}\delta^{bb'}+\delta^{aa'}(t^c_{\bf G})^{bb'}\right)H^{a'b'}~=~H^{ab}+i\alpha^c\left[t^c_{\bf G},H\right]^{ab}~,
\end{equation}
with the representation matrices $(t^c_{\bf G})^{ab}\equiv -if^{abc}$ of the adjoint ${\bf G}$, where $f^{abc}$ are completely antisymmetric structure constants of the gauge group. We normalize the real fields $H^{ab}$ and the gauge fields $G^a_{\mu\nu}\equiv \partial_\mu A^a_\nu-\partial_\nu A^a_\mu+g_Gf^{abc}A^b_\mu A^c_\nu$, along with the gauge coupling $g_G$ and the structure constants $f^{abc}$, in such a way that their kinetic terms have the (standard) form
\begin{equation}
\label{normalizeGH}
{\cal L}~=~\frac{1}{2}(\partial_\mu H^{ab}-ig_GA^c_\mu\left[t^c_{\bf G},H\right]^{ab})(\partial^\mu H^{ab}-ig_GA^{d\mu}\left[t^d_{\bf G},H\right]^{ab})~-~\frac{1}{4}G^a_{\mu\nu}G^{a\mu\nu}
\end{equation}
with the index of the adjoint representation $C({\bf G})\delta^{ab}={\rm tr}(t_{\bf G}^at_{\bf G}^b)=f^{acd}f^{bcd}$ normalized to
\begin{eqnarray}
SU(5):&& C({\bf G})=C({\bf 24})=5~,\label{su5index}\\
SO(10):&& C({\bf G})=C({\bf 45})=8~.\label{so10index}
\end{eqnarray}
The assignment (\ref{so10index}) follows from (\ref{su5index}) if one demands the structure constants of $SO(10)$, restricted to its ``normal'' $SU(5)$ subgroup which we will be interested in later, to coincide with the $SU(5)$ structure constants normalized according to (\ref{su5index}), which itself follows from the standard assignment $C({\bf 5})=1/2$ for the fundamental representation ${\bf 5}$ of $SU(5)$. We use this standard choice $C({\bf N})=1/2$ for the fundamental representation ${\bf N}$ of any unitary group $SU(N)$.

In this notation, the unique gauge (and Lorentz) singlet contraction is
\begin{equation}
HGG~\simeq~H^{ab}G^{a}_{\mu\nu}G^{b\mu\nu}~.
\end{equation}
Writing the dimension-5 operators of interest in this common form for all of the admissible Higgs irreps (\ref{higgsirreps}), together with the normalization (\ref{normalizeGH}) of the kinetic terms of the gauge and Higgs fields, establishes a convention (up to the sign) for the -- a priori arbitrary -- overall factor of the singlet contraction of one Higgs multiplet with two gauge fields. It moreover asserts, as a convention, that the two operators $H_iGG$ and $H_jGG$, even with Higgses $H_i$, $H_j$ in different irreps, are of equal size and strength. The overall sign of the gauge singlet contraction is arbitrary, as it contains the Higgs field $H$ to one power and as there is no invariant way to fix the sign of $H$ through the (quadratic) kinetic term (\ref{normalizeGH}). Consequently, this leaves the overall sign of the Wilson coefficients $c_i$ in (\ref{dim5operators}) open, but allows naturalness assessments based on the absolute values $\vert c_i\vert$.

In (\ref{dim5operators}) we wrote the dimension-5 operators as 
\begin{equation}\label{dim5operatorsapp}
{\cal L} = \sum_i\frac{c_{i}}{M_{Pl}} H_{i}^{ab}\,\frac{1}{4}G^{a}_{\mu\nu}G^{b\mu\nu}~
\end{equation}
(the $H_i$ do not all have to be in distinct representations). The justification for the factor $1/4$ is the same as in the gauge boson kinetic term (\ref{normalizeGH}), as the squared time-derivatives of the spatial gauge fields appear as $2(\partial_0A^a_i)^2$ in the contraction $-G^a_{\mu\nu}G^{a\mu\nu}$, whereas squared time-derivatives of boson fields canonically appear with factors of $1/2$ as kinetic terms in the Lagrangian ${\cal L}$. To avoid the same overcounting, we explicitly put $1/4$ into the operator (\ref{dim5operatorsapp}) as well, noting that this might be a conservative choice (i.e., require a larger value of $\vert c_i\vert$ in order to achieve the same effect). $M_{Pl}$, which we parametrize as $M_{Pl}=1.2\times10^{19}\,{\rm GeV}/\xi$ (cf.~(\ref{suppressionscale})), should be the scale of the physics generating the operators (\ref{dim5operatorsapp}), and we imagine this to be gravitational interactions (see the beginning of Section \ref{sectionSU5setup}). This completes the setup of normalization conventions that satisfy the goal formulated at the beginning of this appendix.

\bigskip

After a Higgs field $H_i$ has acquired a (nonzero) vacuum expectation value (vev) $\langle H_i\rangle$, the effective interaction (\ref{dim5operatorsapp}) contributes a term
\begin{equation}\label{aftervev}
{\cal L}~=~\frac{c_i\langle H^{ab}_i\rangle}{4M_{Pl}}\,G^a_{\mu\nu}G^{b\mu\nu}
\end{equation}
to the Lagrangian, similar to the gauge boson kinetic term in (\ref{normalizeGH}). Our next goal is to look quantitatively at this contribution and its immediate effects.

The contributions (\ref{aftervev}) scale roughly like $\langle H^{ab}_i\rangle/{M_{Pl}}\sim M/{M_{Pl}}$ if $H_i$ assumes its vev at energy scale $M$, and so are completely negligible for $M$ anywhere near or below the electroweak scale. On the other hand, above the electroweak scale the standard model subgroup $G_{321}=SU(3)_C\times SU(2)_L \times U(1)_Y$ of the grand unified group $G$ is unbroken, so the vevs of all scalar fields in the theory, in particular all vevs appearing in (\ref{aftervev}), have to be invariant under $G_{321}$ (this includes zero vev). This requirement on $\langle H^{ab}_i\rangle$ constrains the contributions (\ref{aftervev}) to the gauge kinetic terms. Next we will find, for the cases of $SU(5)$ and $SO(10)$ grand unified gauge groups $G$ and for each of the allowed irreps (\ref{higgsirreps}), all possible $\langle H^{ab}_i\rangle$ that are invariant under the standard model subgroup $G_{321}\subset G$.

The method employed is similar to \cite{Martin:2009ad} (section I), but we do not only want to find relative contributions, but rather also \emph{absolute} values within our normalization conventions above, and we furthermore want to establish a relation between the Higgs vevs and the masses of the superheavy gauge bosons (see later). Let $\phi\equiv(\phi^{ab})$ be a vector, transforming like $\langle H^{ab}\rangle$ as the symmetric product of two adjoint representations ${\bf G}$ of the gauge group $G$, see (\ref{GtimesGtrafo}). Choose\footnote{One way to choose such an explicit basis $t_{\bf G}^a$ of adjoint generators is to choose an explicit basis $t_{\bf F}^a$ for the fundamental representation ${\bf F}$ of $G$, normalized to ${\rm tr}(t_{\bf F}^at_{\bf F}^b)=C({\bf F})\delta^{ab}$, such that the $t_{\bf F}^a$ already obey the desired classification, which may be more easily accomplished than the same classification for the $t_{\bf G}^a$. Here, ${\bf F}={\bf 5}$ or ${\bf 10}$ for $G=SU(5)$ or $SO(10)$, respectively, and $C({\bf 5}_{SU(5)})=1/2$ and $C({\bf 10}_{SO(10)})=1$. Then $(t_{\bf G}^a)^{bc}\equiv-if^{abc}$ with $f^{abc}\equiv-i\,{\rm tr}\left(\left[t^a_{\bf F},t^b_{\bf F}\right]t^c_{\bf F}\right)/C({\bf F})$ have the desired properties.} an explicit basis $\{t_{\bf G}^a\}_a$ of generators of the adjoint representation ${\bf G}$ such that they satisfy the normalization (\ref{su5index}), (\ref{so10index}) and that the generators of the standard model subgroup $G_{123}$ correspond to $a=1,\,\ldots,\,8$ (for the $SU(3)_C$ factor; called ``set $I$'' of the generators), $a=9,\,10,\,11$ ($SU(2)_L$; set $II$) and $a=12$ ($U(1)_Y$; set $III$ or set $III'$, see later); this allows for easy examination of the transformation properties of $\phi$ under the standard model subgroup $G_{321}$. Also, the remaining generators $a=13,\,\ldots,\,d(G)$ may be identified according to their transformation properties under subgroups of the full gauge group $G$, and we need such a naming scheme later to address the components of $\phi^{ab}$ of different superheavy gauge bosons. For our purposes we choose one of the following classifications of generators $t^a$ (applicable to any representation):
\begin{enumerate}
\item For $SU(5)$: The adjoint ${\bf G}={\bf 24}$ of $SU(5)$, under which the generators transform, branches under the standard model subgroup $G_{321}$ into:
\begin{equation}\label{decompSU5}
{\bf 24} ~\stackrel{G_{321}}{\longrightarrow}~ \underbrace{({\bf 8},{\bf 1},0)}_{I} \oplus \underbrace{({\bf 1},{\bf 3},0)}_{II} \oplus \underbrace{({\bf 1},{\bf 1},0)}_{III} \oplus \underbrace{\left(({\bf 3},{\bf 2},-\frac{5}{6}) \oplus {\rm h.c.}\right)}_{IV}~.
\end{equation}
Besides the irreps $I$, $II$ and $III$, there is only one real representation of $G_{321}$ contained in the ${\bf 24}$, and those 12 generators make up set $IV$ of generators.
\item For $SO(10)$: Two different classifications are possible since $G_{321}$ can be embedded in two different ways (normal \cite{Fritzsch:1974nn} or ``flipped'' \cite{Barr:1981qv} embedding) into $SO(10)$ with the correct charge assignments for each standard model family of fermions from a ${\bf 16}$ of $SO(10)$. For either embedding, $SU(5)$ in the following denotes the $SU(5)\subset SO(10)$ subgroup which contains the $SU(3)_C$ and $SU(2)_L$ factors of $G_{321}$, and $U(1)_X$ the Abelian factor such that $SU(5)\times U(1)_X\subset SO(10)$ is a maximal subgroup.
\begin{enumerate}[(i)]
\item normal (Georgi-Glashow-like) embedding $G_{321}\subset SU(5)\subset SO(10)$ \cite{Fritzsch:1974nn}:
\begin{eqnarray}\label{decompSO10GG}
{\bf 45} &\stackrel{SU(5)}{\longrightarrow}& {\bf 24}~~\oplus~~{\bf 1}~~\oplus~~\left({\bf 10}\oplus {\rm h.c.}\right)\nonumber\\
&\stackrel{G_{321}}{\longrightarrow}& \underbrace{({\bf 8},{\bf 1},0)}_{I} \oplus \underbrace{({\bf 1},{\bf 3},0)}_{II} \oplus \underbrace{({\bf 1},{\bf 1},0)}_{III} \oplus \underbrace{\left(({\bf 3},{\bf 2},-\frac{5}{6}) \oplus {\rm h.c.}\right)}_{IV}~~\oplus~~\underbrace{({\bf 1},{\bf 1},0)}_{V}\nonumber\\
&&~~\oplus~~\underbrace{\left(({\bf 3},{\bf 2},\frac{1}{6}) \oplus {\rm h.c.}\right)}_{VI} \oplus \underbrace{\left((\overline{\bf 3},{\bf 1},-\frac{2}{3}) \oplus {\rm h.c.}\right)}_{VII} \oplus \underbrace{\left(({\bf 1},{\bf 1},1) \oplus {\rm h.c.}\right)}_{VIII}~.
\end{eqnarray}
The number of generators in sets $IV$, $V$, $VI$, $VII$, $VIII$ is 12, 1, 12, 6, 2. Sets $I$, $II$, $III$, $IV$ correspond to $I$, $II$, $III$, $IV$ in (\ref{decompSU5}) (this is the Georgi-Glashow embedding). Note, the generators $III$ (generating $U(1)_Y\subset G_{321}$) and $V$ (generating $U(1)_X$) both have the same standard model quantum numbers (both are standard model singlets), whereas all other sets of generators have distinct ones.
\item flipped embedding $G_{321}\subset SU(5)\times U(1)_X\subset SO(10)$ with $G_{321}\nsubseteq SU(5)$ (more precisely, the $U(1)_Y$ factor of $G_{321}$ is not contained in $SU(5)$) \cite{Barr:1981qv}:
\begin{eqnarray}\label{decompSO10flipped}
{\bf 45} &\stackrel{SU(5)\times U(1)_X}{\longrightarrow}& {\bf 24}(0)\oplus{\bf 1}(0)~~\oplus~~\left({\bf 10}(-\sqrt{\frac{2}{3}})\oplus {\rm h.c.}\right)\nonumber\\
&\stackrel{G_{321}}{\longrightarrow}& \underbrace{({\bf 8},{\bf 1},0)}_{I} \oplus \underbrace{({\bf 1},{\bf 3},0)}_{II} \oplus \underbrace{({\bf 1},{\bf 1},0)}_{III'} \oplus \underbrace{\left(({\bf 3},{\bf 2},\frac{1}{6}) \oplus {\rm h.c.}\right)}_{IV}\oplus\underbrace{({\bf 1},{\bf 1},0)}_{V'}\nonumber\\
&&~~\oplus~~\underbrace{\left(({\bf 3},{\bf 2},-\frac{5}{6}) \oplus {\rm h.c.}\right)}_{VI} \oplus \underbrace{\left((\overline{\bf 3},{\bf 1},-\frac{2}{3}) \oplus {\rm h.c.}\right)}_{VII} \oplus \underbrace{\left(({\bf 1},{\bf 1},-1) \oplus {\rm h.c.}\right)}_{VIII}~.
\end{eqnarray}
Again, only the generators $III'$ and $V'$ have the same transformation properties under the standard model. The sets $I$, $II$, $IV$, $VI$, $VII$ and $VIII$ of generators here can be identified with the respective sets in (\ref{decompSO10GG}), whereas the two generators $t^{III'}$ (generating $U(1)_Y\nsubseteq SU(5)$) and $t^{V'}$ here are linear combinations of the generators $t^{III}$ and $t^{V}$ from (\ref{decompSO10GG}),
\begin{equation}\label{relateGGF}
\begin{pmatrix}t^{III'}\\t^{V'}\end{pmatrix}=V\begin{pmatrix}t^{III}\\t^{V}\end{pmatrix}
~~~\text{with}~~~
V=\begin{pmatrix}-\frac{1}{5}\,s^{III}s^{III'}&\frac{2\sqrt{6}}{5}\,s^{V}s^{III'}\\
\frac{2\sqrt{6}}{5}\,s^{III'}s^{V'}&\frac{1}{5}\,s^{V}s^{V'}\end{pmatrix}~,
\end{equation}
where $s^{III},s^{V},s^{III'},s^{V'}=\pm1$ are signs that depend on the choice of signs of the generators $t^{III}$, $t^{V}$, $t^{III'}$ and $t^{V'}$, which are not fixed by our normalization conventions above. $V$ is a real orthogonal matrix, $VV^T={\mathbb I}_2$, and the gauge bosons $F^{III'}_{\mu\nu}$, $F^{V'}_{\mu\nu}$ corresponding to the generators $t^{III'},\,t^{V'}$ are related to the gauge bosons $F^{III}_{\mu\nu}$, $F^{V}_{\mu\nu}$ via $(F^{III'},F^{V'})=(F^{III},F^{V})V^T$.
\end{enumerate}
\end{enumerate}

In this explicit basis, the requirement that $\phi^{ab}$ be invariant under $G_{321}$ then, according to (\ref{GtimesGtrafo}), translates into
\begin{equation}
(t^c_{{\bf G}\otimes{\bf G}})^{(ab)(a'b')}\phi^{a'b'}\equiv\left[t^c_{\bf G},\phi\right]^{ab}=0~~~\text{for all}~\,c=1,\,\ldots,\,12~.
\end{equation}
Solving this system of (linear) equations (along with enforcing symmetry $\phi^{ab}=\phi^{ba}$) yields 4 (resp.~9) linearly independent solutions $\phi_t^{ab}$ ($t=1,\,\ldots,\,4$ resp.~$t=1,\,\ldots,\,9$) in the case of $G=SU(5)$ (resp.~$G=SO(10)$), their linear combinations exhausting all standard model singlet vevs contained in $({\bf G}\otimes{\bf G})_s$. In order to find all possible standard model singlet vevs for each of the Higgs irreps (\ref{higgsirreps}), one has to form linear combinations of the solutions $\phi_t$ that transform in those irreps. To this end, one evaluates the quadratic Casimir operator on each of the linearly independent solutions $\phi_t$, expressing the result again as their linear combination,
\begin{equation}\label{casimironGG}
C_2\phi_t\equiv\sum_{c=1}^{d(G)}\left[t^c_{\bf G},\left[t^c_{\bf G},\phi_t\right]\right]~=~\sum_uc_{ut}\phi_u~,
\end{equation}
which is possible since the restriction of the action of the Casimir operator to an irrep is proportional to the identity on that irrep; Table \ref{C2table} gives the respective constants of proportionality (``quadratic Casimir invariants'') for irreps of several groups of interest here and later. The matrix $(c_{ut})$ is diagonalizable, $\sum_tc_{ut}v^{(w)}_t=C^{(w)}_2v^{(w)}_u$ with linearly independent eigenvectors $v^{(w)}$, $w=1,\,\ldots,\,4\,(9)$ for $G=SU(5)$ ($G=SO(10)$). The 4 (resp.~9) vectors $\Phi_w\equiv\sum_tv_t^{(w)}\phi_t$ then form a basis of the standard model singlet subspace as well, and each $\Phi_w$ transforms in a representation $r$ of $G$ whose quadratic Casimir invariant $C_2(r)$ is equal to the eigenvalue $C_2^{(w)}$. Since, in the $SU(5)$ as well as the $SO(10)$ case, the different irreps in (\ref{higgsirreps}) have distinct quadratic Casimir invariants, and since each irrep only occurs once in the direct sum decomposition (\ref{higgsirreps}) of $({\bf G}\otimes{\bf G})_s$, it is ensured that each $\Phi_w$ actually is contained in an irrep $r$, and $r$ can be read off from the corresponding eigenvalue $C_2^{(w)}$ via Table \ref{C2table}.

\begin{table}[htb]
\begin{center}
\begin{tabular}{l||c|c|c}
\hline
group& $C_2(r_1)$ & $C_2(r_2)$ & $C_2(r_3)$ \\
\hline
\hline
$SU(5)$ & $C_2({\bf 24})=5$ &  $C_2({\bf 75})=8$ &  $C_2({\bf 200})=12$   \\
\hline
$SO(10)$ & $C_2({\bf 54})=10$ &  $C_2({\bf 210})=12$ &  $C_2({\bf 770})=18$   \\
\hline
$SU(4)$ & $C_2({\bf 15})=4$ &  $C_2({\bf 84})=20$ &   \\
\hline
$SU(2)$ & $C_2({\bf 3})=2$ &  $C_2({\bf 5})=6$ &   \\
\hline
\end{tabular}
\end{center}
\caption{Quadratic Casimir invariants $C_2(r)$ for some irreps $r$ of several groups of interest, in the conventions described above. $C_2({\bf 1})=0$ for the singlet ${\bf 1}$ of any group.\label{C2table}}
\end{table}

What one finds, in the $G=SU(5)$ case, is that each of the four irreps ${\bf 1}$, ${\bf 24}$, ${\bf 75}$ and ${\bf 200}$, into which $({\bf 24}\otimes{\bf 24})_s$ decomposes (\ref{higgsirreps}), contains exactly one standard model singlet $\Phi_w$ (the 4 eigenvalues of $c_{ut}$ are nondegenerate: 0, 5, 8, 12). Therefore, in $SU(5)$, specifying the irrep of $H$ determines its (standard model singlet) vev and subsequently its contributions to (\ref{aftervev}) up to an overall factor; in particular, the relative contributions between terms $G^a_{\mu\nu}G^{b\mu\nu}$ in (\ref{aftervev}) are completely determined.

For the $SO(10)$ unified gauge group, however, only the ${\bf 1}$ and ${\bf 54}$ irreps contain exactly one standard model singlet; the ${\bf 210}$ contains three, and the ${\bf 770}$ contains four linearly independent standard model singlets (the 9 eigenvalues of $c_{ut}$ are partly degenerate: 0, 10, $3\times12$, $4\times18$). These statements hold for both the normal (Georgi-Glashow) and the flipped embedding of $G_{321}$ into $SO(10)$, and, in fact, it is found that the set of standard model singlets is the same in either case.

In order to distinguish these linearly independent $G_{321}$-singlets within the ${\bf 210}$ or the ${\bf 770}$ of $SO(10)$, one can specify their transformation properties under subgroups of $SO(10)$ that contain the standard model $G_{321}$ (the normal or the flipped $G_{321}$, respectively). We consider here two such subgroups, namely the two maximal subgroups of $SO(10)$ consistent with the charge assignments for the standard model fermions; these are (a) the extended Georgi-Glashow subgroup $SU(5)\times U(1)_X\subset SO(10)$, and (b) the Pati-Salam subgroup $SU(4)_C\times SU(2)_L\times SU(2)_R$. To achieve this practically, the method of the Casimir operator (\ref{casimironGG}) is employed again; however, the sum over $c$ in (\ref{casimironGG}) is now restricted to the generators of the desired subgroup, or, more precisely, to the generators of each simple factor of the subgroup, in order to find linear combinations of the $\Phi_w$ that transform in irreps of the subgroup (factors). It turns out that either one of the subgroups (a) or (b) can resolve the degeneracy of the vectors $\Phi_w$ transforming within the ${\bf 210}$ or the ${\bf 770}$ of $SO(10)$, see the second column of Tables \ref{SO10entriesSU5U1X} and \ref{SO10entriesSU4CSU2R} for cases (a) and (b), respectively.

So, for each of the irreps $r$ on the right-hand-sides of (\ref{higgsirreps}) in the $SU(5)$ as well as the $SO(10)$ case, we have found a basis $\left\{\Phi_{(r)t}\right\}_t$, such that any standard model singlet $\phi^{ab}$ in the irrep $r$ can be written as a linear combination $\phi^{ab}=\sum_t c_t\Phi^{ab}_{(r)t}$ with real coefficients $c_t$; the index $t$ runs over $t=1$ only, except for $SO(10)$ in the cases of $r={\bf 210}$ (where $t=1,\,2,\,3$) or $r={\bf 770}$ ($t=1,\,2,\,3,\,4$), for which the $\Phi_{(r)t}$ can be given definite transformation properties under (maximal) subgroups of $SO(10)$, as described above. To be specific, we normalize the overall magnitude of each $\Phi_{(r)t}$ such that $\sum_{ab}\Phi_{(r)t}^{ab}\Phi_{(r)t}^{ab}=1$, resulting in $\Phi_{(r)t}^{ab}\Phi_{(r')t'}^{ab}=\delta_{rr'}\delta_{tt'}$. Also, it turns out that $\Phi^{ab}_{(r)t}=0$ whenever $a$ and $b$ belong to two subsets of generators with different transformation properties under $G_{321}$, see (\ref{decompSU5}), (\ref{decompSO10GG}) or (\ref{decompSO10flipped}), and $\Phi^{ab}_{(r)t}$ is proportional to the identity matrix when $a$ and $b$ are restricted to one set of generators. Therefore, nonzero off-diagonal elements $\Phi^{a\neq b}_{(r)t}\neq0$ can possibly occur only in the $SO(10)$ case when $a$ and $b$ correspond to the two degenerate $G_{321}$-singlet generators $III$, $V$ or $III'$, $V'$. In particular, $\Phi^{ab}_{(r)t}$ is diagonal in $a,b$ on the set $a,b=1,\ldots,12$ and is proportional to the identity matrix on each simple factor of the standard model $a,b=1,\ldots,8$ ($SU(3)_C$), $a,b=9,10,11$ ($SU(2)_L$) and $a,b=12$ ($U(1)_Y$); one could therefore arbitrarily, as an invariant convention, for each $r$ and $t$ choose the sign of $\Phi_{(r)t}$ such that, e.g., the nonzero diagonal entry $\Phi^{a=b}_{(r)t}$ with smallest $a=b$ is positive; however, to save some writing, we reverse this convention for the flipped embedding case in rows 5 and 8 of Table \ref{SO10entriesSU4CSU2R}.

We are now ready to write down the matrices $\Phi^{ab}_{(r)t}$, obtained in this way, explicitly. Since, as just described, each matrix $\Phi_{(r)t}$ contains only a few independent entries, they can be written down in an economical way (${\mathbb I}_d$ denotes the identity matrix of size $d\times d$):
\begin{eqnarray}\label{schematicmatrices}\label{schematicmatricesSU5}
\text{for }SU(5):&&\Phi^{ab}=
\left(\begin{smallmatrix}
\Phi^{I}{\mathbb I}_{8}&&&\\
&\Phi^{II}{\mathbb I}_{3}&&\\
&&\Phi^{III}&\\
&&&\Phi^{IV}{\mathbb I}_{12}
\end{smallmatrix}\right)^{ab}~;\\
\text{for }SO(10):&&\Phi^{ab}=
\left(\begin{smallmatrix}
\Phi^{I}{\mathbb I}_{8}\\
&\Phi^{II}{\mathbb I}_{3}\\
&&\Phi^{III}&0&\Phi^{III,V}\\
&&0&\Phi^{IV}{\mathbb I}_{12}&0\\
&&\Phi^{III,V}&0&\Phi^{V}\\
&&&&&\Phi^{VI}{\mathbb I}_{12}\\
&&&&&&\Phi^{VII}{\mathbb I}_{6}\\
&&&&&&&\Phi^{VIII}{\mathbb I}_{2}
\end{smallmatrix}\right)^{ab}~.\label{schematicmatricesSO10}
\end{eqnarray}
For the $SU(5)$ case, the numerical entries of each matrix $\Phi_{(r)t}\equiv\Phi_{(r)}$ (\ref{schematicmatricesSU5}) are listed in Table \ref{SU5entries} (in this case, $t=1$ only, see above). For both the normal and the flipped embedding of $G_{321}\subset SO(10)$, Table \ref{SO10entriesSU5U1X} lists the entries of each $\Phi_{(r)t}$ (\ref{schematicmatricesSO10}) (for the flipped embedding, $\Phi^{III'}_{(r)t}$, $\Phi^{V'}_{(r)t}$ and $\Phi^{III',V'}_{(r)t}$ should be used from Table \ref{SO10entriesSU5U1X} instead of the unprimed $\Phi^{III}_{(r)t}$, etc.), where the $G_{321}$-singlets (labeled by $t$) within each $SO(10)$-irrep $r$ are further classified according to their transformation under the maximal subgroup $SU(5)\times U(1)_X\subset SO(10)$, see above; Table \ref{SO10entriesSU4CSU2R} is similar, but with the $G_{321}$-singlets $\Phi'_{(r)t}$ classified according to their transformation properties under the maximal subgroup $SU(4)_C\times SU(2)_L\times SU(2)_R\subset SO(10)$ (of course, all $G_{321}$-singlets are $SU(2)_L$-singlets, since $SU(2)_L\subset G_{321}$). In all cases, the matrices (\ref{schematicmatricesSO10}) with specified transformation properties under any subgroup are identical for the normal and flipped embedding except for the $\Phi^{III}$, $\Phi^{V}$ and $\Phi^{III,V}$ entries (which correspond to generators $t^{III}$, $t^{V}$ that have the same quantum numbers under $G_{321}$), and these entries are related by
\begin{equation}\label{orthogonalrelationflippednormal}
\begin{pmatrix}\Phi^{III'}&\Phi^{III',V'}\\\Phi^{III',V'}&\Phi^{V'}\end{pmatrix}=
V\begin{pmatrix}\Phi^{III}&\Phi^{III,V}\\\Phi^{III,V}&\Phi^{V}\end{pmatrix}V^T~,
\end{equation}
with the $2\times2$--matrix $V$ from (\ref{relateGGF}).

\begin{table}[htb]
\begin{center}
\begin{tabular}{c||c|c|c|c}
\hline
$r$ & $\Phi^{I}_{(r)}$ & $\Phi^{II}_{(r)}$ & $\Phi^{III}_{(r)}$ & $\Phi^{IV}_{(r)}$ \\
\hline
\hline
${\bf 1}$ & $1/\sqrt{24}$ & $1/\sqrt{24}$ & $1/\sqrt{24}$ & $1/\sqrt{24}$ \\
\hline
${\bf 24}$ & $2/\sqrt{63}$ & $-3/\sqrt{63}$ & $-1/\sqrt{63}$ & $-1/2\sqrt{63}$ \\
\hline
${\bf 75}$ & $1/\sqrt{72}$ & $3/\sqrt{72}$ & $-5/\sqrt{72}$ & $-1/\sqrt{72}$ \\
\hline
${\bf 200}$ & $1/\sqrt{168}$ & $2/\sqrt{168}$ & $10/\sqrt{168}$ & $-2/\sqrt{168}$ \\
\hline
\end{tabular}
\end{center}
\caption{The standard model singlets $\Phi_{(r)}$ in each of the irreps $r$ (\ref{higgsirreps}) of $SU(5)$ in the explicit version (\ref{schematicmatricesSU5}) with the conventions described above.\label{SU5entries}}
\end{table}

\begin{table}[htb]
\begin{center}
\begin{tabular}{c|c||c|c|c|c|c|c|c|c|c||c|c|c}
\hline
$SO(10)$ & $SU(5)\times U(1)_X$ & $\Phi^{I}_{(r)t}$ & $\Phi^{II}_{(r)t}$ & $\Phi^{III}_{(r)t}$ & $\Phi^{IV}_{(r)t}$ & $\Phi^{V}_{(r)t}$ & $\Phi^{III,V}_{(r)t}$ & $\Phi^{VI}_{(r)t}$ & $\Phi^{VII}_{(r)t}$ & $\Phi^{VIII}_{(r)t}$ & $\Phi^{III'}_{(r)t}$ & $\Phi^{V'}_{(r)t}$ & $\Phi^{III',V'}_{(r)t}$ \\
\hline
\hline
$r={\bf 1}$ & $({\bf 1},0)\,,~t=1$ & $\frac{1}{3 \sqrt{5}}$ & $\frac{1}{3 \sqrt{5}}$ & $\frac{1}{3 \sqrt{5}}$ & $\frac{1}{3 \sqrt{5}}$ & $\frac{1}{3 \sqrt{5}}$ & $0$ & $\frac{1}{3 \sqrt{5}}$ & $\frac{1}{3 \sqrt{5}}$ & $\frac{1}{3 \sqrt{5}}$ & $\frac{1}{3 \sqrt{5}}$ & $\frac{1}{3 \sqrt{5}}$ & $0$  \\
\hline
$r={\bf 54}$ & $({\bf 24},0)\,,~t=1$ & $\frac{1}{\sqrt{30}}$ & $-\sqrt{\frac{3}{40}}$ & $-\frac{1}{2 \sqrt{30}}$ & $-\frac{1}{4 \sqrt{30}}$ & $0$ & $\frac{1}{2 \sqrt{5}}$ & $-\frac{1}{4 \sqrt{30}}$ & $\frac{1}{\sqrt{30}}$ & $-\sqrt{\frac{3}{40}}$ & $-\frac{1}{2 \sqrt{30}}$ & $0$ & $\frac{1}{2 \sqrt{5}}$ \\
\hline
$r={\bf 210}$ & $({\bf 1},0)\,,~t=1$ & $\frac{1}{2 \sqrt{15}}$ & $\frac{1}{2 \sqrt{15}}$ & $\frac{1}{2 \sqrt{15}}$ & $\frac{1}{2 \sqrt{15}}$ & $-\frac{2}{\sqrt{15}}$ & $0$ & $-\frac{1}{2 \sqrt{15}}$ & $-\frac{1}{2 \sqrt{15}}$ & $-\frac{1}{2 \sqrt{15}}$ & $-\frac{19}{10 \sqrt{15}}$ & $\frac{2}{5 \sqrt{15}}$ & $-\sqrt{\frac{2}{125}}$ \\
            & $({\bf 24},0)\,,~t=2$ & $\frac{2}{3 \sqrt{15}}$ & $-\frac{1}{\sqrt{15}}$ & $-\frac{1}{3 \sqrt{15}}$ & $-\frac{1}{6 \sqrt{15}}$ & $0$ & $-\frac{1}{\sqrt{10}}$ & $\frac{1}{6 \sqrt{15}}$ & $-\frac{2}{3 \sqrt{15}}$ & $\frac{1}{\sqrt{15}}$ & $\frac{7}{15 \sqrt{15}}$ & $-\frac{4}{5 \sqrt{15}}$ & $-\frac{13}{15 \sqrt{10}}$ \\
            & $({\bf 75},0)\,,~t=3$ & $\frac{1}{6 \sqrt{3}}$ & $\frac{1}{2 \sqrt{3}}$ & $-\frac{5}{6 \sqrt{3}}$ & $-\frac{1}{6 \sqrt{3}}$ & $0$ & $0$ & $\frac{1}{6 \sqrt{3}}$ & $-\frac{1}{6 \sqrt{3}}$ & $-\frac{1}{2 \sqrt{3}}$ & $-\frac{1}{30 \sqrt{3}}$ & $-\frac{4}{5 \sqrt{3}}$ & $\frac{\sqrt{2}}{15}$\\
\hline
$r={\bf 770}$ & $({\bf 1},0)\,,~t=1$ & $\frac{1}{6 \sqrt{10}}$ & $\frac{1}{6 \sqrt{10}}$ & $\frac{1}{6 \sqrt{10}}$ & $\frac{1}{6 \sqrt{10}}$ & $\sqrt{\frac{32}{45}}$ & $0$ & $-\frac{1}{3 \sqrt{10}}$ & $-\frac{1}{3 \sqrt{10}}$ & $-\frac{1}{3 \sqrt{10}}$ & $\frac{77}{30 \sqrt{10}}$ & $\sqrt{\frac{8}{1125}}$ & $\sqrt{\frac{3}{125}}$\\
            & $({\bf 24},0)\,,~t=2$ & $\frac{1}{3 \sqrt{210}}$ & $-\frac{1}{2 \sqrt{210}}$ & $-\frac{1}{6 \sqrt{210}}$ & $-\frac{1}{12 \sqrt{210}}$ & $0$ & $\sqrt{\frac{7}{20}}$ & $\sqrt{\frac{7}{4320}}$ & $-\sqrt{\frac{7}{270}}$ & $\sqrt{\frac{7}{120}}$ & $-\frac{101}{30 \sqrt{210}}$ & $\sqrt{\frac{128}{2625}}$ & $\frac{97}{30 \sqrt{35}}$\\
            & $({\bf 75},0)\,,~t=3$ & $\frac{1}{6 \sqrt{6}}$ & $\frac{1}{2 \sqrt{6}}$ & $-\frac{5}{6 \sqrt{6}}$ & $-\frac{1}{6 \sqrt{6}}$ & $0$ & $0$ & $-\frac{1}{3 \sqrt{6}}$ & $\frac{1}{3 \sqrt{6}}$ & $\frac{1}{\sqrt{6}}$ & $-\frac{1}{30 \sqrt{6}}$ & $-\sqrt{\frac{8}{75}}$ & $\frac{1}{15}$\\
            & $({\bf 200},0)\,,~t=4$ & $\frac{1}{2 \sqrt{42}}$ & $\frac{1}{\sqrt{42}}$ & $\frac{5}{\sqrt{42}}$ & $-\frac{1}{\sqrt{42}}$ & $0$ & $0$ & $0$ & $0$ & $0$ & $\frac{1}{5 \sqrt{42}}$ & $\sqrt{\frac{96}{175}}$ & $-\frac{2}{5 \sqrt{7}}$\\
\hline
\end{tabular}
\end{center}
\caption{The standard model singlets $\Phi_{(r)t}$ in each of the irreps $r$ (\ref{higgsirreps}) of $SO(10)$, classified according to their transformation properties under the $SU(5)\times U(1)_X\subset SO(10)$ maximal subgroup (second column; note, all standard model singlets $\Phi_{(r)t}$ also happen to be $U(1)_X$-singlets), in the explicit version (\ref{schematicmatricesSO10}) with the conventions described above. The entries of the matrices $\Phi_{(r)t}$ (\ref{schematicmatricesSO10}) agree between the normal and the flipped embedding of $G_{321}\subset SO(10)$, except for the $\Phi^{III}_{(r)t}$, $\Phi^{V}_{(r)t}$ and $\Phi^{III,V}_{(r)t}$ entries, where the primed ones ($\Phi^{III'}_{(r)t}$, etc., see last three columns) should be used instead for the flipped embedding, cf.~also (\ref{orthogonalrelationflippednormal}). (For both the normal and the flipped embedding, the $SU(5)$ here denotes the $SU(5)\subset SO(10)$ containing $SU(3)_C\times SU(2)_L$.)\label{SO10entriesSU5U1X}}
\end{table}

\begin{table}[htb]
\begin{center}
\begin{tabular}{c|c||c|c|c|c|c|c|c|c|c||c|c|c}
\hline
$SO(10)$ & $SU(4)_C\times SU(2)_R$ & $\Phi'^{I}_{(r)t}$ & $\Phi'^{II}_{(r)t}$ & $\Phi'^{III}_{(r)t}$ & $\Phi'^{IV}_{(r)t}$ & $\Phi'^{V}_{(r)t}$ & $\Phi'^{III,V}_{(r)t}$ & $\Phi'^{VI}_{(r)t}$ & $\Phi'^{VII}_{(r)t}$ & $\Phi'^{VIII}_{(r)t}$ & $\Phi'^{III'}_{(r)t}$ & $\Phi'^{V'}_{(r)t}$ & $\Phi'^{III',V'}_{(r)t}$ \\
\hline
\hline
$r={\bf 1}$ & $({\bf 1},{\bf 1})\,,~t=1$ & $\frac{1}{3 \sqrt{5}}$ & $\frac{1}{3 \sqrt{5}}$ & $\frac{1}{3\sqrt{5}}$ & $\frac{1}{3 \sqrt{5}}$ & $\frac{1}{3 \sqrt{5}}$ & $0$ & $\frac{1}{3 \sqrt{5}}$ & $\frac{1}{3 \sqrt{5}}$ & $\frac{1}{3\sqrt{5}}$ & $\frac{1}{3 \sqrt{5}}$ & $\frac{1}{3 \sqrt{5}}$ & $0$ \\
\hline
$r={\bf 54}$ & $({\bf 1},{\bf 1})\,,~t=1$ & $\frac{1}{\sqrt{30}}$ & $-\sqrt{\frac{3}{40}}$ & $-\frac{1}{2 \sqrt{30}}$ & $-\frac{1}{4 \sqrt{30}}$ & $0$ & $\frac{1}{2 \sqrt{5}}$ & $-\frac{1}{4 \sqrt{30}}$ & $\frac{1}{\sqrt{30}}$ & $-\sqrt{\frac{3}{40}}$ & $-\frac{1}{2 \sqrt{30}}$ & $0$ & $\frac{1}{2 \sqrt{5}}$ \\
\hline
$r={\bf 210}$ & $({\bf 1},{\bf 1})\,,~t=1$ & $0$ & $\frac{1}{\sqrt{6}}$ & $-\sqrt{\frac{3}{50}}$ & $0$ & $-\sqrt{\frac{2}{75}}$ & $\frac{1}{5}$ & $0$ & $0$ & $-\frac{1}{\sqrt{6}}$ & $-\sqrt{\frac{3}{50}}$ & $-\sqrt{\frac{2}{75}}$ & $\frac{1}{5}$ \\
            & $({\bf 15},{\bf 1})\,,~t=2$ & $\frac{1}{3 \sqrt{2}}$ & $0$ & $-\frac{2 \sqrt{2}}{15}$ & $0$ & $-\frac{\sqrt{2}}{5}$ & $-\frac{2}{5 \sqrt{3}}$ & $0$ & $-\frac{1}{3 \sqrt{2}}$ & $0$ & $-\frac{2 \sqrt{2}}{15}$ & $-\frac{\sqrt{2}}{5}$ & $-\frac{2}{5 \sqrt{3}}$ \\
            & $({\bf 15},{\bf 3})\,,~t=3$ & $0$ & $0$ & $\frac{2}{5}$ & $\frac{1}{6}$ & $-\frac{2}{5}$ & $\frac{1}{5 \sqrt{6}}$ & $-\frac{1}{6}$ & $0$ & $0$ & $-\frac{2}{5}$ & $\frac{2}{5}$ & $-\frac{1}{5 \sqrt{6}}$ \\
\hline
$r={\bf 770}$ & $({\bf 1},{\bf 1})\,,~t=1$ & $\frac{1}{3 \sqrt{10}}$ & $\sqrt{\frac{5}{72}}$ & $\frac{19}{30 \sqrt{10}}$ & $-\sqrt{\frac{5}{288}}$ & $\sqrt{\frac{32}{1125}}$ & $-\sqrt{\frac{3}{500}}$ & $-\sqrt{\frac{5}{288}}$ & $\frac{1}{3 \sqrt{10}}$ & $\sqrt{\frac{5}{72}}$ & $\frac{19}{30 \sqrt{10}}$ & $\sqrt{\frac{32}{1125}}$ & $-\sqrt{\frac{3}{500}}$ \\
            & $({\bf 1},{\bf 5})\,,~t=2$ & $0$ & $0$ & $\frac{\sqrt{6}}{5}$ & $0$ & $\sqrt{\frac{8}{75}}$ & $-\frac{2}{5}$ & $0$ & $0$ & $-\frac{1}{\sqrt{6}}$ & $\frac{\sqrt{6}}{5}$ & $\sqrt{\frac{8}{75}}$ & $-\frac{2}{5}$ \\
            & $({\bf 15},{\bf 3})\,,~t=3$ & $0$ & $0$ & $\frac{2 \sqrt{2}}{5}$ & $-\frac{1}{6 \sqrt{2}}$ & $-\frac{2 \sqrt{2}}{5}$ & $\frac{1}{5 \sqrt{3}}$ & $\frac{1}{6 \sqrt{2}}$ & $0$ & $0$ & $-\frac{2 \sqrt{2}}{5}$ & $\frac{2 \sqrt{2}}{5}$ & $-\frac{1}{5 \sqrt{3}}$ \\
            & $({\bf 84},{\bf 1})\,,~t=4$ & $\frac{1}{6 \sqrt{10}}$ & $0$ & $\sqrt{\frac{128}{1125}}$ & $0$ & $\sqrt{\frac{32}{125}}$ & $\frac{8}{5 \sqrt{15}}$ & $0$ & $-\sqrt{\frac{2}{45}}$ & $0$ & $\sqrt{\frac{128}{1125}}$ & $\sqrt{\frac{32}{125}}$ & $\frac{8}{5 \sqrt{15}}$ \\
\hline
\end{tabular}
\end{center}
\caption{Same as Table \ref{SO10entriesSU5U1X}, but with the standard model singlets $\Phi'_{(r)t}$ classified according to their transformation properties under the $SU(4)_C\times SU(2)_L\times SU(2)_R\subset SO(10)$ maximal subgroup (second column; all entries are $SU(2)_L$ singlets, since $SU(2)_L\subset G_{321}$). (The $SU(4)_C$ here denotes the $SU(4)$ factor in $SU(4)\times SU(2)\times SU(2)\subset SO(10)$ with $SU(4)\supset SU(3)_C$, where one of the two $SU(2)$ factors is $SU(2)_L$.)\label{SO10entriesSU4CSU2R}}
\end{table}

Furthermore, for each $SO(10)$-irrep $r$ separately, the vevs $\Phi_{(r)t}$, classified according to $SU(5)\times U(1)_X$ (see Table \ref{SO10entriesSU5U1X}), and the vevs $\Phi'_{(r)t}$, classified according to $SU(4)_C\times SU(2)_R$ (Table \ref{SO10entriesSU4CSU2R}), are linear combinations of each other, in fact, they are related by orthogonal transformations; when their overall signs are chosen as shown in the tables, then, for both the normal and the flipped embedding, the relation between the $\Phi_{(r)t}$ and the $\Phi'_{(r)t}$ is:
\begin{eqnarray}\label{orthogrel}
\Phi'_{({\bf 1})1}=\Phi_{({\bf 1})1}~,~~
\Phi'_{({\bf 54})1}=\Phi_{({\bf 54})1}~,~~
\begin{pmatrix}\Phi'_{({\bf 210})1}\\\Phi'_{({\bf 210})2}\\\Phi'_{({\bf 210})3}\end{pmatrix}
=\begin{pmatrix}\frac{1}{\sqrt{10}} & -\sqrt{\frac{2}{5}} & \frac{1}{\sqrt{2}} \\ \sqrt{\frac{3}{10}} & \sqrt{\frac{8}{15}} & \frac{1}{\sqrt{6}} \\ \sqrt{\frac{3}{5}} & -\frac{1}{\sqrt{15}} & -\frac{1}{\sqrt{3}}\end{pmatrix}
\begin{pmatrix}\Phi_{({\bf 210})1}\\\Phi_{({\bf 210})2}\\\Phi_{({\bf 210})3}\end{pmatrix}~,\nonumber\\
\begin{pmatrix}\Phi'_{({\bf 770})1}\\\Phi'_{({\bf 770})2}\\\Phi'_{({\bf 770})3}\\\Phi'_{({\bf 770})4}\end{pmatrix}
=\begin{pmatrix}\frac{1}{5} & -\sqrt{\frac{3}{175}} & \sqrt{\frac{3}{5}} & \sqrt{\frac{12}{35}} \\ \sqrt{\frac{3}{20}} & -\frac{4}{\sqrt{35}} & -\frac{1}{2} & \frac{1}{\sqrt{7}} \\ -\frac{3}{2 \sqrt{5}} & \frac{2}{\sqrt{105}} & -\frac{1}{2 \sqrt{3}} & \sqrt{\frac{3}{7}} \\ \frac{3}{5} & \frac{16}{5 \sqrt{21}} & -\frac{1}{\sqrt{15}} & \sqrt{\frac{3}{35}}\end{pmatrix}
\begin{pmatrix}\Phi_{({\bf 770})1}\\\Phi_{({\bf 770})2}\\\Phi_{({\bf 770})3}\\\Phi_{({\bf 770})4}\end{pmatrix}~.
\end{eqnarray}

By classifying the standard model singlets $\Phi_{(r)t}$, in the $SO(10)$ case, into irreps under (maximal) subgroups of $SO(10)$ as in Tables \ref{SO10entriesSU5U1X} and \ref{SO10entriesSU4CSU2R}, we do not mean to imply a grand unified symmetry breaking scenario where $SO(10)$ is broken to the standard model $G_{321}$ necessarily via some intermediate gauge group, although this classification is well suited for such a scenario, see the end of section \ref{sectionSO10}. Moreover, such a classification in particular can serve as a parametrization for standard model singlets $\langle H_i^{ab}\rangle$ (\ref{aftervev}), transforming in irreps $r_i$, in terms the of the basis vectors $\Phi_{(i)t}\equiv\Phi_{(r_i)t}$:
\begin{eqnarray}
\langle H_i^{ab}\rangle~\equiv~\sum_t v_{(i)t}\Phi_{(i)t}^{ab}~\equiv~v_i\phi_i^{ab}\label{parametrizeHiggsVEV}
\end{eqnarray}
with real scalars $v_{(i)t}$, $v_i$, and where, e.g.~for $r_i={\bf210}$,
\begin{equation}
\phi^{ab}_{i}~=~\Phi^{ab}_{({\bf 210})1}\cos\theta_1+\Phi^{ab}_{({\bf 210})2}\sin\theta_1\cos\theta_2+\Phi^{ab}_{({\bf 210})3}\sin\theta_1\sin\theta_2~,
\end{equation}
such that $\phi^{ab}_{i}\phi^{ab}_{i}=1$ and $\sum_t v_{(i)t}^2=v_i^2$ for all $i$; then the kinetic terms of the $v_i$ and $v_{(i)t}$ field degrees of freedom, before assuming vevs, are canonical (\ref{normalizeGH}) ${\cal L}=\frac{1}{2}\,(\partial_\mu H_i^{ab}+\ldots)(\partial^\mu H_i^{ab}+\ldots)=\frac{1}{2}\,(\partial_\mu v_i)^2+\ldots$. (The two parametrizations in (\ref{parametrizeHiggsVEV}) differ only if $H_i$ is a ${\bf 210}$ or ${\bf 770}$ of $SO(10)$, since in all other cases the sum merely runs over $t=1$.)

\bigskip

We assume, for definiteness and simplicity (see Sections \ref{sectionSU5setup} and \ref{sectionSO10}), one-step breaking of the grand unified gauge group $G$ to the standard model $G_{321}$ at the unification scale $M_X$. Below $M_X$, all Higgs multiplets $H_i$ responsible for grand unified symmetry breaking assume nonzero vevs (\ref{parametrizeHiggsVEV}), and in particular give masses to the non--$G_{321}$ gauge bosons (henceforth called ``superheavy'' gauge bosons, although not all of them can get mass from the Higgses (\ref{higgsirreps}), see below), see (\ref{normalizeGH}):
\begin{equation}
{\cal L}=\frac{1}{2}g_G^2A^a_\mu A^{b\mu}\sum_i{\rm Tr}\left(-\left[t^a_{\bf G},\langle H_i\rangle\right]\left[t^b_{\bf G},\langle H_i\rangle\right]\right)\equiv\frac{1}{2}m^2_{ab}A^a_\mu A^{b\mu}~;
\end{equation}
each Higgs $H_i$ contributes independently to the gauge boson squared mass matrix $m^2_{ab}$. For an $SU(5)$ grand unified group, each of the 12 superheavy gauge bosons acquires equal mass:
\begin{equation}\label{SU5gbmass}
SU(5):~~m^2_{ab}=\sum_i \frac{C_2(r_i)}{12}g_G^2v_i^2\,\delta_{ab}\equiv M_{\rm gb}^{2}\,\delta_{ab}~~~\text{for}~a,b=13,\,\ldots,24~,
\end{equation}
where $C_2(r_i)$ is the quadratic Casimir invariant of irrep $r_i$ (cf.~Table \ref{C2table} for the irreps (\ref{higgsirreps})), and $v_i$ the vev of $H_i$ (cf.~(\ref{parametrizeHiggsVEV}) and below). In $SO(10)$, the situation is more complicated: Higgs multiplets in any of the irreps (\ref{higgsirreps}) of interest for the dimension-5 operators (\ref{dim5operatorsapp}) fail to give mass to all of the superheavy gauge bosons. In particular, the non--$G_{321}$ gauge boson $G^V_{\mu\nu}$ (or $G^{V'}_{\mu\nu}$ for the flipped embedding) belonging to the set $V$ (or $V'$; see (\ref{decompSO10GG}) or (\ref{decompSO10flipped})) of generators is always left massless by Higgses in irreps (\ref{higgsirreps}), provided only that their vevs are $G_{321}$-singlets, cf.~also the second column of Table \ref{SO10entriesSU5U1X}. Also, some of the other superheavy gauge bosons might remain massless, and, at any rate, do not receive equal masses. Higgses $H_i$ in representations $r_i$ other than those on the right hand side of (\ref{higgsirreps}) are needed to give mass to all of the 33 non--$G_{321}$ gauge bosons in the $SO(10)$ case, and they cannot occur in dimension-5 operators (\ref{dim5operatorsapp}). Independent of the individual directions $\phi_i^{ab}$ (\ref{parametrizeHiggsVEV}) of the vevs $\langle H_i\rangle=v_i\phi_i$, the averaged superheavy gauge boson squared mass is:
\begin{equation}\label{SO10gbmass}
SO(10):~~\overline{m^2}_{ab}=\sum_i \frac{C_2(r_i)}{33}g_G^2v_i^2\,\delta_{ab}\equiv M_{\rm gb}^{2}\,\delta_{ab}~~~\text{for}~a,b=13,\,\ldots,45~,
\end{equation}
where the sum runs over all Higgs multiplets at the grand unification scale, whether or not they occur in (\ref{dim5operatorsapp}).

Grand unification asserts that the masses of the superheavy gauge bosons are related to the scale $M_X$ of gauge coupling unification; namely, their masses have to be somewhere around $M_X$, such that the renormalization group equations of the standard model apply below the scale $M_X$, at which the gauge couplings unify and above which the unified gauge coupling evolves according to the $\beta$--function of the unified gauge theory. For definiteness in obtaining numerical values, we assert for the analysis in the main text, that the superheavy gauge boson masses (or the averaged superheavy gauge boson mass (\ref{SO10gbmass}) in the $SO(10)$ case) have to exactly equal the unification scale:
\begin{equation}
M_{\rm gb}~=~M_X~.
\end{equation}
In the $SU(5)$ case, any of the Higgses in (\ref{higgsirreps}) can give mass to all of the superheavy gauge bosons; so we assert, again for definiteness (see Sections \ref{sectionSU5setup} and \ref{sectionSU5results}), that no Higgs multiplets in irreps other than (\ref{higgsirreps}) are present in the theory (nor, for that matter, in the sum (\ref{SU5gbmass})). In the $SO(10)$ analysis, however, we assert that the Higgs multiplets in irreps (\ref{higgsirreps}) account for some fraction $1/f$, e.g.~half, of the average gauge boson squared mass:
\begin{eqnarray}\label{assertMX}
M_X~=~M_{\rm gb}~&=&~\sqrt{\frac{f}{d({G})-12}\,\sum_{\{i\,\vert\,r_i\text{~in~}(\ref{higgsirreps})\}}C_2(r_i)\,g_G^2v_i^2}\\
~&=&~g_Gv~\sqrt{\frac{f}{d({G})-12}\,\sum_{\{i\,\vert\,r_i\text{~in~}(\ref{higgsirreps})\}}C_2(r_i)\,x_i^2}~,\label{assertMXwithratios}
\end{eqnarray}
with $f=1$ (resp.~$f=2$) for $SU(5)$ (resp.~$SO(10)$), and $d({G})-12=12$ (resp.~$=33$) the number of superheavy gauge bosons. The form (\ref{assertMXwithratios}) arises when the vevs $v_i=x_i v$ are to obey some given ratio $x_1:x_2:\ldots\,$, with an open overall scale $v$, which will be useful later.

\bigskip

Now, plugging the Higgs vevs (\ref{parametrizeHiggsVEV}) back into the Lagrangian (\ref{aftervev}) and adding the gauge boson kinetic term (\ref{normalizeGH}), one obtains:
\begin{eqnarray}
{\cal L}&=&-\frac{1}{4}G^a_{\mu\nu}G^{a\mu\nu}+\sum_i\frac{c_i}{4M_{Pl}}\,\sum_t v_{(i)t}\Phi_{(i)t}^{ab}\,G^a_{\mu\nu}G^{b\mu\nu}\nonumber\\
&=&-\frac{1}{4}\left(1-\sum_i\frac{c_i}{M_{Pl}}\sum_t v_{(i)t}\Phi_{(i)t}^{I}\right)F^a_{\mu\nu}F^{a\mu\nu}_{SU(3)}\nonumber\\
&&-\frac{1}{4}\left(1-\sum_i\frac{c_i}{M_{Pl}}\sum_t v_{(i)t}\Phi_{(i)t}^{II}\right)F^a_{\mu\nu}F^{a\mu\nu}_{SU(2)}\nonumber\\
&&-\frac{1}{4}\left(1-\sum_i\frac{c_i}{M_{Pl}}\sum_t v_{(i)t}\Phi_{(i)t}^{III}\right)F_{\mu\nu}F^{\mu\nu}_{U(1)}~+~\ldots~\nonumber\\
&\equiv&-\frac{1}{4}(1+\epsilon_3)F^a_{\mu\nu}F^{a\mu\nu}_{SU(3)}-\frac{1}{4}(1+\epsilon_2)F^a_{\mu\nu}F^{a\mu\nu}_{SU(2)}-\frac{1}{4}(1+\epsilon_1)F_{\mu\nu}F^{\mu\nu}_{U(1)}~+~\ldots~~.\label{withepsilonsapp}
\end{eqnarray}
Here, we have used the forms (\ref{schematicmatricesSU5}), (\ref{schematicmatricesSO10}) of the standard model singlets $\Phi_{(i)t}$ on the subspace $I$, $II$, $III$ (or $III'$) of $G_{321}$-generators and have omitted the superheavy gauge bosons, assuming the Lagrangian (\ref{withepsilonsapp}) to be valid below the unification scale $M_X$ where they have acquired mass and are integrated out; the last line defines what we mean by $\epsilon_s$, $s=1,2,3$, namely the corrections to the gauge boson kinetic terms originating from the Higgs multiplets in the dimension-5 operators assuming nonzero vevs (\ref{aftervev}):
\begin{equation}\label{defepsilonAPP}
\epsilon_s=\sum_i\frac{c_i}{M_{Pl}}\sum_t v_{(i)t}\delta_{s}^{(i)t}~~(\text{for}~s=1,2,3),~\text{where}\,~\delta_1^{(i)t}\equiv-\Phi_{(i)t}^{III}\,,~\,\delta_2^{(i)t}\equiv-\Phi_{(i)t}^{II}\,,~\,\delta_3^{(i)t}\equiv-\Phi_{(i)t}^{I}~;
\end{equation}
the $\delta_s^{(i)t}$ can be read off from Tables \ref{SU5entries}, \ref{SO10entriesSU5U1X} and \ref{SO10entriesSU4CSU2R} for the $SU(5)$ and the $SO(10)$ cases (use $\delta_1^{(i)t}\equiv-\Phi_{(i)t}^{III'}$ for the flipped embedding $G_{321}\subset SO(10)$), and Table \ref{deltaSU5table} in the main text gives them for $SU(5)$ explicitly (in this case, $t=1$ only).

As shown in Section \ref{sectionSU5setup}, after rescaling $F^{\mu\nu}_{(s)}\to(1+\epsilon_s)^{1/2}F^{\mu\nu}_{(s)}$ and $g_s\to(1+\epsilon_s)^{-1/2}g_s$ to the observed low-energy gauge field strengths and gauge couplings by amounts which are different for each of the standard model gauge group factors $s=1,2,3$, the condition for gauge coupling unification at the scale $M_X$ into the unified group reads, in terms of the observed (running) gauge couplings $\alpha_s=\alpha_s(\mu)=g_s(\mu)^2/4\pi$ of the theory below $M_X$:
\begin{equation}\label{unificationconditionAPP}
(1+\epsilon_1)\alpha_1(M_X)=(1+\epsilon_2)\alpha_2(M_X)=(1+\epsilon_3)\alpha_3(M_X)~=~\frac{g_G^2}{4\pi}\equiv\alpha_G~,
\end{equation}
where $g_G\equiv g_G(M_X)$ is the gauge coupling of the unified theory at the scale $\mu=M_X$. The functions $\alpha_s(\mu)$ are fixed through their low-energy measurements (e.g., at the scale $\mu=m_Z$ of the $Z$ mass) and their renormalization group evolution; in the non-supersymmetric standard model to one-loop order with one standard model Higgs doublet:
\begin{equation}\label{runningcouplingsAPP}
\frac{1}{\alpha_s(\mu)}=\frac{1}{\alpha_s(m_Z)}-\frac{b_s}{2\pi}\ln\frac{\mu}{m_Z}
\end{equation}
with $\beta$--function coefficients
\begin{equation}\label{betafunctioncoeffsAPP}
b_1=\frac{41}{10}\,,~b_2=-\frac{19}{6}\,,~b_3=-7~,
\end{equation}
and initial values \cite{Amsler:2008zzb}
\begin{equation}\label{initialvaluesAPP}
\alpha_1(m_Z)=0.016887\pm0.00040\,,~\alpha_2(m_Z)=0.03322\pm0.00025\,,~\alpha_3(m_Z)=0.1176\pm0.005
\end{equation}
for the couplings at the scale $m_Z=91.1876\,{\rm GeV}$. For two-loop evolution, which we use in Section \ref{SU5oneoperatorsection} for comparison to one-loop, see \cite{Jones:1982}.

\bigskip

Under the perspective of this Appendix, unification is the numerical requirement that the unification scale $M_X$, the unified gauge coupling $g_G$, the Wilson coefficients $c_i$ and the Higgs vevs $v^2_i=\sum_t v^2_{(i)t}$ satisfy equations (\ref{unificationconditionAPP}), (\ref{defepsilonAPP}) and (\ref{assertMX}) simultaneously, with the given running coupling functions $\alpha_s(\mu)$ of the standard model, e.g.~(\ref{runningcouplingsAPP})--(\ref{initialvaluesAPP}) to one loop. In the main text, we further demand that the values of these quantities are ``natural'', so that physically sensible unification can be claimed, and we exhibit such models.

We outline here a procedure to solve these equations simultaneously, after having chosen the unified gauge group ($SU(5)$ or $SO(10)$) and the Higgs content $H_i$ (the multiplets that may occur in the dimension-5 operators (\ref{dim5operatorsapp})): Fix any ratio $x_1:x_2:\ldots\,$ between the Higgs vevs $v_i=x_iv$ (without loss of generality, $\sum_i x_i^2=1$ with, e.g., spherical coordinates); for Higgs multiplets $H_i$ in a ${\bf 210}$ or a ${\bf 770}$ of $SO(10)$, also fix the direction of the vev $\langle H_i\rangle$ by fixing the ratio $y_{(i)1}:y_{(i)2}:\ldots\,$ between the $v_{(i)t}=y_{(i)t} v_i$ (with $\sum_t y^2_{(i)t}=1$), see (\ref{parametrizeHiggsVEV}). Also fix the ratio $z_1:z_2:\ldots\,$ between the Wilson coefficients $c_i=z_i c$ (with $\sum_i z_i^2=1$). With these inputs, the \emph{ratio} between the $\epsilon_s$ ($s=1,2,3$) in (\ref{defepsilonAPP}) is completely fixed:
\begin{equation}\label{howtosolve}
\epsilon_s=\frac{cv}{M_{Pl}}\,\sum_i z_i\sum_t x_iy_{(i)t}\delta_s^{(i)t}~.
\end{equation}
A key observation is now that any such given ratio $\epsilon_1:\epsilon_2:\epsilon_3$, along with the given functions $\alpha_s(\mu)$, determines $M_X$ and the values $\epsilon_s$ uniquely by solving the two left equalities in (\ref{unificationconditionAPP}); this can be seen analytically to one-loop order when (\ref{runningcouplingsAPP}) is plugged back into (\ref{unificationconditionAPP}) (the linear system of equations for $\,\ln M_X$ has nonvanishing determinant for almost any given ratio $\epsilon_1:\epsilon_2:\epsilon_3$), but also holds at two-loop order. Then, the unified coupling $\alpha_G=g_G^2/4\pi$ is determined by (\ref{unificationconditionAPP}) as well, and so is the overall scale $v$ of the vevs via (\ref{assertMXwithratios}). Finally, with a choice (\ref{suppressionscale}) of the Planck scale $M_{Pl}$, the required Wilson coefficients $c_i=z_i c$ can be computed from (\ref{howtosolve}). Note, that this last step is the only place where the choice of the Planck scale $M_{Pl}=1.2\times10^{19}\,{\rm GeV}/\xi$ comes in: the smaller a Planck scale one chooses, the smaller the Wilson coefficients $c_i$ have to be in order to achieve unification (at the same $x_i$, $y_{(i)t}$, $z_i$); in fact, the necessary Wilson coefficients are inversely proportional to the choice of $\xi$ in (\ref{suppressionscale}).

\end{document}